COMPUTER NETWORK DEFENSE THROUGH

RADIAL WAVE FUNCTIONS

by

Ian J. Malloy

A Thesis Submitted to the Faculty of

Utica College

December, 2015

in Partial Fulfillment of the Requirements for the Degree of

Master of Science in
Cybersecurity

**Abstract**

The purpose of this research was to synthesize basic and fundamental findings in quantum computing, as applied to the attack and defense of conventional computer networks. The concept focuses on uses of radio waves as a shield for, and attack against traditional computers. A logic bomb is analogous to a landmine in a computer network, and if one was to implement it as non-trivial mitigation, it will aid computer network defense. As has been seen in kinetic warfare, the use of landmines has been devastating to geopolitical regions in that they are severely difficult for a civilian to avoid triggering given the unknown position of a landmine. Thus, the importance of understanding a logic bomb is relevant and has corollaries to quantum mechanics as well. The research synthesizes quantum logic phase shifts in certain respects using the Dynamic Data Exchange protocol in software written for this work, as well as a C-NOT gate applied to a virtual quantum circuit environment by implementing a Quantum Fourier Transform. The research focus applies the principles of coherence and entanglement from quantum physics, the concept of expert systems in artificial intelligence, principles of prime number based cryptography with trapdoor functions, and modeling radio wave propagation against an event from unknown parameters. This comes as a program relying on the artificial intelligence concept of an expert system in conjunction with trigger events for a trapdoor function relying on infinite recursion, as well as system mechanics for elliptic curve cryptography along orbital angular momenta. Here trapdoor both denotes the form of cipher, as well as the implied relationship to logic bombs.

## Acknowledgements

This work is in dedication to the memory of Dr. Daniel Lee Swets.

# Table of Contents

# List of Illustrative Materials

Computer Network Defense Through

Radial Wave Functions

      The purpose of this research was to synthesize computing security principles using logic and mathematics to design mitigation capabilities for cyber security algorithms to mitigate quantum computer threats. The research focused on reverse engineering logic bombs for use as a defensive tool within the purview of context-specific activation and execution. The foundational theory of quantum Turing machine's applications to the engineering of a cyber-security system was novel. Using an analysis of quantum logic phase shifts researched by Q.A. Turchette, C.J. Hood, W. Lange, H. Mabuchi, and H.J. Kimble as funded by the National Science Foundation and Office of Naval Research this mitigation system uses their findings (1995, p. 4714). The aim of applying the phase shifts were masking methods to simulate logic bomb attacks, and as a foundation for mitigation strategies. The validity of the research in the proof of concept used a quantum circuit virtual environment, as well as Wolfram Mathematica software.

      While normal computers work with quantum mechanics, until 2012 no traditional computer or conventional communication electronics were functioning quantum communication systems. Researchers at The Cambridge Research Laboratory in conjunction with Toshiba Research have succeeded in the extraction of information using quantum communication but did so using "ordinary telecom fibres [sic] transmitting data traffic" (Physics World, 2012, para. 7). This is contrary to some modern opinions of 2015 which state that algorithmic implementations of quantum computing in traditional computing environments are neither possible nor quantum computing.

      While a quantum computer uses either an electron or photon to perform the calculations it runs, conventional computers operate the same in the sense that they too use electrons, or

electricity and magnetism. In sharp contrast to traditional computers, which run series of (0) and(1), a quantum computer can calculate using a state called superposition where it is both (0) and (1). A conventional computer using the zeroes and ones creates "bits" or "bytes" while the quantum version is a "qubit" or even "qudit."

The security of quantum communication comes, then, from "entanglement" and "coherence." Arthur Pittenger is professor *emeritus* form the University of Maryland, Baltimore College and alumnus of Stanford where he earned his B.S., M.S. and Ph.D., and explains when two photons transition to entanglement, they act the same at exactly the same time with no respect to distance, as long as their entanglement is coherent (Pittenger, 1999, p. 5). Since the Earth has a magnetic field, quantum communication with photons is difficult to keep in coherence. The magnetism from the Earth's core creates a "noise" which leads to the entanglement becoming "decoherent."

**A Brief History of Computation**

In Gregory Chaitin's chapter, titled *A Random Walk in Arithmetic*, Chaitin, a pioneer in algorithmic information theory proposed that "Turing showed that there is no set of instructions that you can give the computer, no algorithm that will decide if a program will ever halt" (1991, p. 199, para. 1). What this means is that a Turing machine, or the first concept of any computer, could not stop running a program with an instruction to stop at a given moment. The machine enters "recursion," or an infinite loop.

The Turing machine operates using a tape, with either (1) or (0) written on the tape. A head on the machine reads the tape, and at a certain bit, it either stops or changes the bit. The problem, called the "Halting Problem" is when a universal Turing machine, or one that can calculate any given problem, is given a random problem and random instructions, therefore becoming a machine running on probability.

C. Monroe, D.M. Meekhof, B.E. King, W.M. Itano, and D.J. Wineland with funding from the Office of Naval Research and United States Army discussed the relevance of processing speed for classes of problems in their research (1995, p. 4714). The team of scientists under the direction of Monroe found that, "The most dramatic example is an algorithm presented by Shor showing that a quantum computer should be able to factor large numbers very efficiently" (1995, p. 4714, para. 2). Their conclusion, the impact on the security of conventional computers, is that halting affects classical computers when quantum computers can operate on the same problem set within seconds and halt where a normal computer cannot (Monroe, et al., 1995, p. 4714).

**Classical and Quantum Turing Machines**

David Deutsch developed a quantum Turing machine (QTM) in 1985 based upon a Turing machine (1985, pp. 97-117). The article, as published by the proceedings of the Royal Society, argued that an implicit physical assertion existed within the Church-Turing thesis (Deutsch, 1985, p. 97). Deutsch extended this further by explicitly stating, "…every finitely realizable physical system can be perfectly simulated by a universal model computing machine operating by finite means" (1985, p. 97, para. 1). This contradicted the traditional understanding of the Halting Problem since any Turing machine that is universal, said to be "Turing Complete" cannot operate within Deutsch's boundaries using finite means. In order for a finitely operating machine to calculate all physically realizable systems, it requires the use of quantum mechanics. Normal, everyday computers do use quantum mechanics, but not the way a quantum computer uses quantum mechanics.

Pittenger explains quantum mechanics by sharing a story about Albert Einstein's frustration with it since probability is the only way to begin an understanding of quantum mechanics (1999, p. 5). According to Pittenger, Einstein decidedly thought quantum mechanics was fundamentally incomplete (1999, p. 5). This led to Einstein's statement that, "God doesn't play dice." Meaning physical reality *should* be entirely predictable with high levels of accuracy, not degrees of confidence.

The fundamental part of quantum mechanics for computing are the gates researched by Monroe and their fellow team members (1995, pp. 4714-4717). The gate is of an XOR form using negation on qubits (Monroe, et al., 1995, p. 4714). An XOR logical operation done by traditional computers takes a bit value, and either accepts or rejects in response based on the value of the bit. The concept behind an XOR gate is based on the "exclusion" of an "either/or" logical operation. Applied to computation, the XOR function is true only if (*A*) and (*B*) are either true or false, and if only one is true, the XOR is true.

The demonstration of a quantum inference based on XOR involved the use of a Controlled-NOT gate from quantum logic principles (Monroe, et al., 1995, p. 4714). The interpretation is an entangled state of two spin-up particles $\langle 11| \uparrow\uparrow\rangle$ shifting phases into a superposition of spin-up and spin-down $\langle 01| \downarrow\uparrow\rangle$. Monroe with their colleagues succinctly cover the principle, "[the target qubit]...is flipped depending on the state of the 'control' qubit" (Monroe, et al., 1995, p. 4714, para. 3). This means that either a photon (packet of light) or an electron (a negatively charged particle separated from an atom) can be targeted and manipulated to do what normal computers do, only better. Equation 1 shows the ground state of a qubit.

$$|00\rangle \tag{1}$$

Equations 2-6 from the work of Monroe and the research team demonstrate C-NOT gates, which are a quantum logic operation (Monroe, et al., 1995, p. 4715):

$$|0\rangle|\downarrow\rangle \rightarrow |0\rangle|\downarrow\rangle \tag{2}$$

$$|0\rangle|\uparrow\rangle \rightarrow |0\rangle|\uparrow\rangle \tag{3}$$

$$|1\rangle|\downarrow\rangle \rightarrow |1\rangle|\uparrow\rangle \tag{4}$$

$$|1\rangle|\uparrow\rangle \rightarrow |1\rangle|\downarrow\rangle \tag{5}$$

$$\{|11\rangle|\uparrow\uparrow\rangle \rightarrow |01\rangle|\downarrow\uparrow\rangle\} \tag{6}$$

**The Logic of Explosions Covered**

Insofar as the ability to program a logic bomb, the following pseudo-code written by Goodrich & Tamassio (2011, p. 177, Figure 4.2) in their introductory text to cyber security serves the purpose well of showing how they operate.

```
{
if
(trigger-condition = TRUE)
activate bomb
;
}
```

The application of a logic bomb as defense does require reclassification according to Goodrich and Tamassio as stated in their text (2011, p. 178). They stated that if a logic bomb is not malicious, such as the "Y2K bug," the operation is not a logic bomb (Goodrich & Tamassio, 2011, p. 178). While the Y2K bug did not do damage, logic bomb that are malicious still need to execute specific functions. A logic bomb is a program that performs a malicious action when activated (Goodrich & Tamassio, 2011, p. 178). The non-intentional, non-malicious "bugs" or flaws in code do not qualify as logic bombs according to Goodrich and Tamassio, since the structure of logic bombs is not present (2011, p. 178). Goodrich and Tamassio offer the following functions as necessary requirements for classifying algorithms as a logic bomb: 1) Trigger, 2) Target, 3) Access, 4) Arm, 5) Launch Payload, and 6) Cover Process.

These six subroutines form the case of Tim Lloyd as acting against Omega Engineering Corporation in 1996 (Goodrich & Tamassio, 2011, p. 178). While the use of logic bombs make for good Hollywood movies such as in the first *Jurassic Park* film where a programmer is able to steal embryos from his employer by using a logic bomb attack, the Omega case shows the actual threats posed by such malicious activity (Goodrich & Tamassio, 2011, p. 178). Thankfully, the evidence gathered by the U.S. Secret Service proved beyond a reasonable doubt that Lloyd intentionally programmed a logic bomb within a server under his administration. Lloyd was guilty of a cybercrime by violating the Computer Fraud and Abuse Act according to the United States' law. He is guilty of exceeding authorized use of a computer network.

**Understanding the Quantum Problem**

The purpose of quantum mechanics is to inform us of the correct method to construct operators corresponding to physical quantities of which we aim to measure, according to Harvard alum and Fulbright lecturer Jay Anderson (2002, p. 3). The meaning behind the methods of quantum mechanics is to explain, with strong accuracy, how the physical universe operates according to fundamental laws. Part of the intersection between mathematics and physics are the concepts of a vector and scalar. A scalar is one-dimensional, while a vector is a physical value that follows in a direction and has magnitude.

The roots of quantum mechanics itself, comes from two different viewpoints signifying two analogous mathematical approaches to eigenvalue problems, where an eigenvalue problem leads to an eigenfunction, which may be an eigenvector (Anderson, 2002, p. 3). A scalar is the dot product of two vectors (Anderson, 2002, p. 106). An eigenfunction may be an eigenvalue, but are also a set of functions, which are independent of one another that solve differential equations. P.R. Wallace, a pioneer in bringing theoretical physics to Canadian Universities in the second half of the 20$^{th}$ century, treats differential equations as explanations of the transformation of a function upon a variable, "…since the transforms of derivatives of a function are proportional to the transform of the function itself" (Wallace, 1984, p. 199). Quantum computing operates using gates, which according to Turchette and the team, requires entanglement and coherence (1995, p. 4714). When two particles enter entanglement, it means they communicate regardless of distance, thus the eigenvector related is different from a classical understanding of vectors and may be an eigenfunction. A function in general is an equation that manipulates variables or equations. Results reported by *Network World* from a speech given at a Black Hat conference in 2013 discuss how quantum computing may very well result in a "cryptoapocalypse" given the functions a quantum computer can perform (Nelson, 2015, para.1). The conference proceeding discusses an acute point irrespective of entanglement, but reliant on the factoring capabilities of quantum computers.

    The article from *Network World* was a summary of principles from a Black Hat conference on implications about quantum computing and encryption (Nelson, 2015, para. 2). Because of the advanced computing speed that quantum computers possess, some fear current cryptography will not survive in a quantum-computing world (Nelson, 2015, para. 4). According to Nelson, "Quantum computing already promises to make existing cryptography easily breakable" (2015, Security Implications, para. 16 ). The *Network World* reporter wrote that the

threat is so severe, that the National Security Agency of the United States is working towards "quantum resistant algorithms in the not-too-distant future" (Nelson, 2015, NSA, para. 17).

The goal, accorded by the National Security Agency (NSA) is to, "…provide cost effective security against a potential quantum computer" (2015, Background, para.3). The relevance of this desire on part of the NSA is critical given a "space race" currently underway according to a dated physics organization report (Physics World, 2013). Quantum communication satellites for quantum communication channels are in development by researchers Thomas Jennewein and Brenden Higgins working out of Cambridge and a department of Toshiba, which began in 2013 (Physics World, 2013, para. 1). The reasoning behind this is to avoid the interference from the Earth's magnetic field. Active investigations into the potential of using space as a vehicle for quantum communication attenuate noise from Earth's magnetic field and thus enable more coherent quantum communications (Physics World, 2013, para. 7). In accordance with the need for algorithmic mitigation of quantum cracking, the resilience of cyber systems is an integral component.

The resilience of computer networks and cyber systems begin with active preservation and continuation of operations throughout attacks, according to Allan Friedman and P.W. Singer's research (2014, p. 170). Friedman and Singer are active members of the Brookings Institute as specialists in information and cyber security. The researchers go on to state it is worth noting that resiliency cannot (or should not) be separated from the human element (Singer & Friedman, 2014, p. 172). Resilience will serve to maintain operations essential to everyday, civilian life, such as National Critical Infrastructure Systems (NCIS).

Literature Review

Jason Andress and Steve Winterfeld, two researchers whom hold several acclaimed certifications in the field of cyber security, reported that logical operations are capable of physical effects (Andress & Winterfeld, 2014, p. 139). Insofar as physical attacks can change logical operations, logical attacks can, in turn, deny or degrade physical systems (Andress & Winterfeld, 2014, p. 139). Even though hardware is physical, it both operates using, and enables logical executions (Andress & Winterfeld, 2014, p. 139). The process of enacting an attack using logic is the foundation of malicious software, "malware." While there are reports nearly weekly of breaches, not all use malware. Resilience requires humans because some attacks use social engineering, where one person manipulates another to reveal data to use in an attack. In cases that do not use social engineering, malware can infect and destroy targeted systems.

**Turing Machines**

The fundamental definition of a Turing machine (TM) is the transition function, ($\delta$) since this dictates and describes how the machine processes information. This is a universally accepted fact explained by an MIT computer science professor Michael Sipser (1997, p. 35). Related to this is the analogy of a TM in quantum mechanics through applying the concept of the halting problem with probability (Chaitin, 1991, p. 199). The quantum randomness is not only an attribute of physics, but also pure mathematics (Chaitin, 1991, p. 196). Despite randomness, however, quantum theory does give a more "faithful reproduction" of qualitative characteristics of experience than any preceding theory according to Nobel Prize winning physicist Percy Bridgman (1964, p. 111). While wave mechanics presents characteristics of error, this does not speak to the accuracy of wave mechanics (Bridgman, 1964, p. 112). These concepts of wave mechanics, probability, and TM can expand into a result of a multi-tape structure for a quantum Turing machine (QTM):

$$[\delta : (Q \text{ X } \Gamma^K)] \rightarrow [(Q \text{ X } \Gamma^K) \text{ X } \{L, R\}^K] \tag{7}$$

Where *(K)* is the number of tapes in Equation 7 (Sipser, 1997, p. 136). Transition functions for non-deterministic TM's have the form of Equation 8 according to Dr. Sipser (Sipser, 1997, p. 138):

$$[\delta : (Q \text{ X } \Gamma)] \rightarrow [p(Q \text{ X } \Gamma) \text{ X } \{L, R\}] \tag{8}$$

Where computation of a non-deterministic TM is a tree, of which the branches correspond to *(p)* events of the machine from Equation 8 by modifying Sipser's model (1997, p. 138). A contention in structuring a QTM lies between the aforementioned chance of error in quantum mechanics and the fact that a true TM is not liable to error, which was a finding of mathematician Cole Kleene, who developed recursion theory and worked with Alan Turing.

The infinite memory of a TM in general allows for the computation of a value for($\mathbb{N}$), or the natural numbers, as arguments for some (*a*) values or null values of (*a*) (Kleene, 1967, p. 260). If, for (*a*), $[f(a) = C]$ then it is demonstrated that the TM computes the function of (*a*), or $f(a)$ and $f(a)$ is computable (Kleene, 1967, p. 260). The positive and converse of the Church-Turing thesis is that every Turing computable function is intuitively computable and these two senses are equivalent (Kleene, 1967, p. 232).

The methodology to begin engineering a QTM hinges on several key factors according to David Deutsch, the first person to develop a QTM. Deutsch's advancement of the Church-Turing thesis for the development of his QTM proposes standards that need to be satisfied for an acceptable QTM (Deutsch, 1985, p. 105). Alan Turing illustrated that no set of instructions, nor algorithm given to a computer, will determine if a program will "halt" (Sipser, 1997, p. 160). In regards to general information theory, the probabilistic relation between a signal and its source is a probability *(p)* of being in the *(i^th)* state, the entropy of which per symbol for the machine as a source is:

$$H = \sum_i P_i H_i \qquad (9)$$

Equation 9 expresses the bits per symbol as a function of the sum of the probability multiplied by the entropy. The entropy, $H_i$ of state $(i)$ is in accordance with Equation 10:

$$H_i = -\sum_{i=1}^{n} p_i(j) \log(p_i j) \qquad (10)$$

Equation 10 leads to the following summation equation:

$$H_i = -\sum_j p_i(j) \log p_i(j) \qquad (11)$$

The function $(\log x)$ is a function of the cause of effect "$x$" where the actual calculation is a relationship between a radical and exponent relative to "$x$" values. Robert Ash, professor *emeritus* in mathematics from the University of Illinois, wrote in 1965 how the relationships between a signal and noise, or interference is relative to entropy in that the higher the level of entropy, the higher the uncertainty is concerning a signal (1990, p. 24). Gaussian distributions relate to general information theory as well given the randomness of prime numbers (Ash, 1990, p. 240).

**Alice, Bob, and Quantum Security**

Gaussian distributions compared to any distribution with a given variance, have added uncertainty (Ash, 1990, p. 240). Gaussian distributions are capable of being both discrete and continuous in nature as communication channels (Ash, 1990, p. 240). Ash proves, that along with use of a Hilbert space, "It is possible to give an *explicit* procedure for constructing codes for the time-continuous Gaussian channel which maintain any transmission rate up to half the channel capacity with an arbitrarily small probability of error" (1990, p. 254, para. 1).

Ian Chant of the IEEE Spectrum website, a leading source on engineering breakthroughs reports that in 2014 researchers were able to transmit 32 Gigabits of data over the air using orbital angular momentum (para.1, 2014). Another writer for IEEE Spectrum, Alexander Hellemans reported in 2012 that what Chant stated was met with strong skepticism by experts (para. 12, 2012). What the experts self-reported to IEEE in 2012 was done during the time that Toshiba and Cambridge were using quantum communication with traditional telecommunication fibers (Physics World, 2012, para. 7).

The time it takes to break an encryption algorithm, or cipher is an important component of the strength of the algorithm. Sipser defines this state as a probabilistic TM, where $p(TM)$ is a one-way permutation in polynomial time. Quantum computers operate in exponential time, but with $p(TM)$ the probability that a variable $(w)$ does not equal $(n)$ is a function of the probability that the TM is in state $(n^{-k} = n/k)$, where $(n)$ is any random number (1997, pp. 375,377). Prime numbers and randomness are both intertwining and critical to functional cryptographic defense against quantum cracking. Cracking encryption was the birth of the modern computer.

The story of Alan Turing is well known in cryptography, used to discuss the impact "cracking" cryptography can have by using the German commanding of their U-boats during World War II as a case study (Sipser, 1997, p. 372). "Alice and Bob," fictional characters, can communicate securely using quantum communication which Jim Alves-Foss, director of Center for Secure and Dependable Systems discusses (n.d., p. 2.1). Alves-Foss explains how quantum communication works by addressing the methods of extracting the message from signals without noise (n.d., p. 2.1). The signal returns random values to anyone who reads the signal incorrectly (Alves-Foss, n.d., p. 2.1). Essentially, any tampering is easily discoverable by Alice and Bob (Alves-Foss, n.d., p. 2.1). The principle of quantum mechanics, known as entanglement, guarantees that Alice can tell Bob the key values irrespective of any time component (Alves-Foss, n.d., p. 2.2).

While quantum computing poses a risk, the additional research from 2013 by Daniel Genkin, Adi Shamir and Eran Tromer working out of Tel Aviv in Israel, whom now work for the Technical Institute of Israel illustrates the conventional threat posed against the RSA encryption algorithms. Their work uses acoustic noises, or radio signals, from various laptop models to crack an RSA algorithm and extract actionable data should they have chosen any attack they simulated (2013, p. 5). They used parabolic microphones, which is a curved sensor and then reflected the sensor from the same curve to allow a longer distance of an attack vector (Genkin, Shamir, & Tromer, 2013, p. 5). While they admit it *only* takes an hour for this attack scenario, the contrast of how long a quantum computer would take is of significantly less time.

**Quantum Satisfiability**

Schrodinger created the first method to solving eigenvalue problems using differential expressions (Anderson, 2002, p. 3). Heisenberg created the other, using operators relying on algebraic methods and matrices (Anderson, 2002, p. 3). Edwin Taylor has received awards for his contributions in teaching physics as well as teaching at the Massachusetts Institute of Technology, and co-authored a text on physics with A.P. French (who worked on the Manhattan Project during World War II). They treat a particle traversing a plane in accordance with the perpendicular relationship that results.

If a particle with angular momentum traverses a plane classically, then it points perpendicular to the plane (French & Taylor, 1978, p. 460). This then means that the angular momentum lies along the z-axis, classically speaking (French & Taylor, 1978, p. 460). To draw an analogy to this in quantum mechanics, the particle described by a 2-D wave function $\psi(x, y)$ is the resulting Eigen-function of a specific operation. If $(q)$ is a prime or a power of a prime, the elements $(0,1,\dots,q-1)$ create a *finite* field under addition and multiplication where

$$S = \left[\frac{q^n}{\sum_{i=0}^{e}(\hat{i})(q-1)}\right] \tag{12}$$

If Equation 12 holds true, and $(q)$ is prime, any Abelian group of *q-ary* sequences is considered a vector space over the field $mod(q)$, therefore only the group structure is needed when $(q = 2)$ (Ash, 1990, p. 96). The following conditions are required for Abelian groups of which the direct sum is a cyclic group, also of which the sub-groups are pure and a principal ideal ring. Equations 13-16 refine the conditions for an Abelian group. Principal ideal rings are commutative, possess abnormal division of properties, and factor uniquely into prime elements.

$$\left(\frac{G}{H}\right), where\ (y_i mod(H)) \tag{13}$$

$$\left(G = (H \oplus K)\right) \tag{14}$$

$$\left[[(t^*) = \sum a_i y_i] \uparrow \left[[Z - \sum a_i x_i] \oplus [0 \in \frac{G}{H}]\right]\right] \tag{15}$$

$$[\sum a_i x_i \in K] \tag{16}$$

Let $(K)$ be a sub-group of $(G)$ generated by $(x_i)$, then the resulting summation becomes an element of set $(K)$ such that set $(Z)$ is an element of the addition of sets $(H)$ and $(K)$. Equation 17 is the concluding space as an element of the resulting group. This is a membership of the sum between the sub-group and divisor of the parent group.

$$\therefore [Z \in (H + K)] \tag{17}$$

At the point where set $(H)$ intersects set $(K)$ at the zero of the functions, there is a set $(W)$ that is equal to Equation 15 and $(W)$ is an element of set $(H)$. Given these resulting conditions, the point $(y_i)$ when equal to infinity then approaches $(a_i)$ at its zero value. This only occurs if $(y_i)$ does not equal infinity, and requires some $(n_i)$ not equal to infinity. The final conditions for a pure sub-group within these boundaries is that $(a_i)$ is a multiple of $(n_i)$, and also that $(t)$ is a multiple of $(n_i)$. With these satisfied requirements, the subgroup calculated is then pure.

Principle rings as a rule have only one prime element. The use of primes within cyber security serve a function in key systems for cryptography, and within the purview of quantum mechanics and information theory, primes remain essential to cryptography also. With a ring of *p-adic* primes, expressed as $(Rp^*)$, the ring must contain some variable, which is in relation to the single prime element shown by Equation 18.

$$(a_0 + a_1 p \dots a_n p_n^n) \tag{18}$$

The satisfiability of any algebra, $(A)$, of two languages $(L)$ and $(L')$ when $(L)$ is a proper sub-group of $(L')$ means that some interpretation of the languages, $(I)$ is an interpretation of $(L)$ on set $\{S\}$. The interpretation $(I')$ of $(L')$ on set $\{S'\}$ such that each interpretation and its respective language are in a proportional relationship to one another as a proper sub-group to the set. Equation 19 is the mathematics of the proportions of sets that are in union with the languages and interpretations:

$$[(I' \bowtie L') \subseteq \{S\}'] \equiv [\{S\} \cup (I \bowtie L)] \tag{19}$$

Furthermore, with each interpretation and related language, there are functions specific to each set where each function is an element and unique to each language. These components to the language allow for the designation of another set, $(A)$ wherein this new set and the interpretation, $(I)$ result in a multiplicative union of set $(I')$ with both languages as well as $\{S\}$ (see Equation 20).

$$[(A = (A, I) \rightarrow (I' \otimes (S, L, L')] \tag{20}$$

The final requirement to prove satisfiability of a logical and mathematical system for engineering algorithms is that some $(F)$ must be computable starting at the value$(A_n)$. With the foundation of decidability and the structure of the group established, it is necessary to explain the purposes of these calculations under the purview of isomorphic groups. The simplest and most basic of any algebraic system is a group, so described by mathematician Charles Pinter who earned the prestigious *State Doctorate* at the University of Paris. In geometry isomorphism has several types, with the simplest in turn being similarity and congruence (Pinter, 1990, p. 90). Two geometric figures are congruent if there is a plane motion where the motion makes one figure coincide with the other (Pinter, 1990, p. 90).

The figures are similar if a transformation of fixed proportion affects the length in a given ratio (Pinter, 1990, p. 90). If $(G)$ is a group and$(a \in G)$, it is feasible to say every element of $(G)$ is a power of$(a)$. Therefore all elements of $(G)$ are a power of $(a)$ and nothing else (Pinter, 1990, p. 93). Pinter demonstrates the conditions for a generator element of some set, which produces a cyclic group. The group itself contains only elements of which are powers of the cyclic generator shown in Equation 21.

$$[G = \{\hat{a} : (n \in Z)\}] \tag{21}$$

The proper expression of generator elements of a cyclic group is Equation 22.

$$(G = \langle a \rangle) \tag{22}$$

If there is some group,$(H)$ that is homomorphic with$(G)$, then a function$(f)$, transforms $(G)$ into $(H)$ (Pinter, 1990, p. 137). When a vector, $(x \neq 0)$ is an eigenvector of$(A)$, if $(A)$ carries $(x)$ into a collinear vector, the value lambda$(\lambda)$, is the eigenvalue of operator$(A)$, corresponding to the eigenvector $(x)$ writes expert Georgi Shilov, who conducted pioneering work in generalized functions and functional analysis (Linear Algebra, 1977, p. 108). A vector space over a defined field is a set with addition and multiplication defined (Pinter, 1990, p. 283). This is vector addition and scalar multiplication (Pinter, 1990, p. 283).

**Hilbert space.** The necessity of establishing definitions of ($\mathfrak{H}$) as a Hilbert space is critical to completing the formalization of the groundwork necessary to engineer a quantum computing system (Shilov, 1974, Elementary Functional Analysis). Shilov provides the following requirements to satisfy conditions of a Hilbert space as accepted in mathematics and computer science (see Equations 23-26) (Elementary Functional Analysis, 1974). ($\mathfrak{H}$), said to be a Hilbert space, if for every pair of vectors a defined real number or scalar product satisfies four axioms expressed as Equations 23-26.

$$(x, y) > 0 \text{ if } x \neq 0 \tag{23}$$

$$(x, y) = (x, y) \text{ for all } (x, y) \text{ in the Hilbert space} \tag{24}$$

$$(\alpha x, y) = (x, y) \text{ for all } (x, y) \text{ in the Hilbert space} \tag{25}$$

$$\forall (x, y, z)[(x + y, z) = (x, z) + (y, z)] \tag{26}$$

A complex system or any system with imaginary components involved is a complex linear ($\mathfrak{H}$) if all vector pairs are a scalar product of the conjunction of vector pairs that satisfy Equations 23-26 (Shilov, Elementary Functional Analysis, 1974, p. 63).

The relation between imaginary numbers to physical values is difficult to address well. Henry E. Kyburg Jr. offers a solid structure such as this from which an extrapolation for engineering a QTM can be based upon. Kyburg, drawing from his expertise in both philosophy (having received the Butler Medal for Philosophy), and his strong knowledge of scientific principles structures bases of a dynamic system involving boundary conditions. The boundary conditions of his formal logic descriptions of physical sciences, and the change it undergoes, is a system of both probability and actuality (Kyburg Jr., 1968, p. 222).

The increase in internal energy or the heat input to a system and the work performed by the system create the foundation of any classical mechanics (Kyburg Jr., 1968, p. 239). As explained by Kyburg, for any system the definition of the change in entropy expresses the change in transition from one state to another (1968, p. 241). The second law of thermodynamics has probabilistic character, according to Kyburg's understanding, stating entropy always increases, and proceeds in directions of increasing probabilities (Kyburg Jr., 1968, p. 241).

The need for a Hilbert space comes from the need for a Hilbert system to establish valid Turing machines, explains Mordechai Ben-Ari, a recipient of the 2004 ACM SIGCSE Award for his contributions in explaining computation and mathematical logic. Hilbert systems are deductive for single formulas (Ben-Ari, 2001, p. 48). ($\mathfrak{H}$), a deductive system with a tri-axiomatic scheme, has one rule of inference (Ben-Ari, 2001, p. 48). To verify the existence of a vector space over a real field, the following Equations 27-29 must hold according to Ben-Ari. These are the rules of inference.

$$\vdash (A \to (B \to A)) \tag{27}$$

$$\vdash \left((A \to (B \to C)) \to ((A \to B) \to (A \to C))\right) \tag{28}$$

$$\vdash \left((\neg B \to \neg A) \to (A \to B)\right) \tag{29}$$

To verify the existence of a vector space over a real field, the following holds. Where each positive integer, $(k)$ for $(R^k)$ when $(R^k)$ is the set of all ordered *k-tuples*, values of some $(x)$ are real numbers and coordinates of a set $(X)$ and the elements of $(R^k)$ are vectors. A vector space $(R^k)$ can then exist over a real field.

A function of two real variables is said to be harmonic on a domain if the second partial derivatives exist and are continuous within that domain. Every point of that domain must satisfy the respective partial derivative as a zero of Laplace's equation. The Bochner-Weil theorem states a continuous function on the same domain is positive definite only if there is a variable, which has a value on that domain, where Equation 30 is satisfiable.

$$p(x) = \int_{G'}^{G} \langle x, x' \rangle d\langle x' \rangle \leftarrow (\forall x: x \in G) \tag{30}$$

With respect to Equation 30 (see page 18), $(G)$ represents the domain as discussed to satisfy the positive definite identity required. The boundary conditions of the now complete formulization unify with the principle of synchronous, concurrent algorithms (SCA) as engineered by B. Thompson, J. Tucker, and J. Zucker, who are experts in mathematics, computation, and complexity theory. An SCA as created by the researchers working with Thompson describe it as a process based on modules within networks and channels, which compute and communicate data in parallel, synchronized by a global clock with discrete time (2009, p. 1386). SCA are useful given the applications it has to analyze and develop coupled-map lattices, based upon discrete time, with some discrete and continuous space (Thompson, et al., 2009, p. 1386).

**Fields, physics, and light.** The Lorentz transformation is of great significance to the conceptualization of any QTM. The Lorentz transformations listed (Equations 31-34) have far-reaching applicability:

$$x' = \left( \frac{x - vt}{\sqrt{1 - \frac{v^2}{c^2}}} \right) \tag{31}$$

$$y' = y \tag{32}$$

$$z' = z \tag{33}$$

$$t' = \left( \frac{t - \frac{v^2}{c^2} * x}{\sqrt{1 - \frac{v^2}{c^2}}} \right) \tag{34}$$

When the transmission of a particle of light travels along the positive *x-axis*, the light-stimulus adheres to Equation 35 according to the world-renowned physicist Albert Einstein (1961, p. 38):

$$(x = ct) \tag{35}$$

The foundation of particles and fields are vectors of $(x)$ according to their respective norms, which result in groups. Asim Orhan Barut, a former professor at Syracuse University and former co-director of the Institute for Theoretical Physics explains the relationship between electrodynamics and fields as a function between Lorentz transformations and relativistic quantum theories. Barut formulizes these relationships into categories within group theory, such that Equations 36-38 hold true (1980, p. 8).

$$x^2 > 0 \tag{36}$$

$$x^2 = 0 \tag{37}$$

$$x^2 < 0 \tag{38}$$

With respect to system behavior, Equation 36 describes time-like vectors, Equation 37 represents light-like or null vectors, and finally space-like vectors are within Equation 38. Complex Lorentz spaces have direct applications to relativistic quantum theories, according to Barut (1980, p. 11). By adding imaginary four-vector $(iy)$ to every real four-vector, $(x)$ the following derivation may be calculated as Equation 39.

$$(\xi = x - iy) \tag{39}$$

Two generalizations of Lorentz space are an ordinary complex vector space with real but non-positive norms having a scalar product. Equations 40-43 solidify the application of Lorentz transformations using the principles of Equations 36-39 to produce the eigenfunction as Equations 40-43 .

$$\xi n = \overline{\xi n} \tag{40}$$

$$[(\xi(\alpha n + \beta n_2)) = (\bar{\alpha}\xi n_1 + \bar{\beta} n_2 \xi)] \tag{41}$$

$$[((\alpha \xi_1 + \beta \xi_2)n) = (\alpha \xi_1 n + \beta \xi_2 n)] \tag{42}$$

$$[\xi^2 = x^2 + y^2] \tag{43}$$

Wave mechanics are the core to any quantum mechanics system given the duality of light, which take on forms of both particles and waves. A center-of-mass momentum has the general form of Equations 44 and 45.

$$(E_1, +P_1) \tag{44}$$

$$(E_2, -P_2) \tag{45}$$

The limit of motion for the relativistic particles is a derivative of the wave function such that the derivative and wave are at the zero-value of the expression as Equation 46.

$$\rho \ddot{\psi} - Y \left(\frac{\partial^2 \psi}{\partial x^2}\right) = 0 \tag{46}$$

The dot notation above the wave function, $(\ddot{\psi})$ symbolizes the second derivative with respect to time. The wave function follows the propagation mechanic of Equation 46, where $(v)$ is the constant velocity and $(\rho)$ has a physical correlation to mass per unit length. Therefore, Equation 47 is the vector relative to velocity, and $(Y)$ is Young's modulus.

$$v = \left(\frac{\sqrt{Y}}{\rho}\right) \tag{47}$$

Where the amplitude of the wave function is $(\psi(x, y))$. This specific function coincides with the Euler-Lagrange Equations. Most importantly, the dynamics of the field as an action principle is Equation 48.

$$\delta \int \mathcal{L} d^3 x dt = 0 \tag{48}$$

## Methodology

The first step to develop security relative to quantum information was to demonstrate relationships between quantum systems and classical computers. Using findings from quantum computing research along with previously developed systems of quantum logic reverse engineering was necessary. The methodology involved using expert system pseudo-code with modification to develop new algorithms called *QUINE*. Lastly, elements of quantum logic phase shifts were the basis of construction for a virtual quantum circuit built with QuIDE. The elements of phase shifts as implemented were inverse transition shifts, as well as the principle of entanglement and coherence. The initial stages of this research were engineering the mathematical foundation for implementation of algorithms in accordance with Equations 18-20 (see page 15) which are formulizations for a computationally satisfiable system.

The software for modeling the quantum circuit is open-source and free to use, and the result of graduate work by Joanna Patrzyk and Bartlomiej Patrzyk for their MSc. degrees. This work from the CGW Conference in 2014, "QuIDE" is on the official website at "http://www.quide.eu/." SWI-Prolog is a robust development environment for several languages. The official website for downloading SWI-Prolog is at www.swi-prolog.org.

### Software Environments

A Windows 32-bit operating system running on a desktop computer served as the environment for the research. The QuIDE quantum computer interface allowed change to the source code such that a user is able to embed C# scripts or functions to the virtual environment. By adjusting the Wolfram Mathematica functions into MatLab syntax and saving the MatLab code, it was not possible to import them into the QuIDE environment given the incompatible version of Mathematica used. Since consumers choose the Windows 32 bit OS environment given the widespread use cases this is the reason for applying it to the research. Thus, while the calculations may not be repeatable by any reader, the ability to cut and paste the code both written for this research as well as the C# script of QuIDE is able to be implemented by a wider range of readers. Furthermore, the private sector makes wide use of Windows OS based servers, and the NCIS of the United States operates using Windows OS environments, thus use-cases of *QUINE* are more applicable to these forms of networks (Malloy, 2015).

The modeling of equations that are results from the methodology were graphed using Wolfram Mathematica version 8, Student Edition. Mathematica allows each graph to progress over time. SWI-Prolog is a programming language known for the applications it has for building artificial intelligence, and is the Edinburgh standard of the Prolog language family. The version of SWI-Prolog used is 7.3.9. Buffers written in prolog are executable using the Windows terminal by running a prolog shell from the buffer, typing "cmd.exe" within the shell and starting the shell, and finally saving the buffer while the shell is running using the ".bat" file extension. Mathematica is a product of the Wolfram Research company founded by Stephen Wolfram in 1987 (Wolfram Technology, 2015, p. About Wolfram Research). The company goal of Wolfram is to develop technology tools so computation is more powerful with each product release (Wolfram Technology, 2015, p. About Wolfram Research). The primary use of Mathematica was to generate the graphs of equations to aid in explanations of functions. In addition, the graphs generated by Mathematica show the behavior of the equations.

**Transcendental Complex Identities**

Friedrich Gauss was a brilliant mathematician who worked to calculate the occurrence of twin-prime numbers, where twin primes are two, consecutive odd numbers where the division of each results in both itself and the number one. Daniel Shanks whose work is available from the American Mathematical Society explored the relationship of the variable $(m)$ such that the algorithm $(m^2 + 1)$ is a prime number (1959). The equations specifically engineered for the purpose of this research demonstrate the potential for use of radio wave mechanics to mitigate quantum computer threats. Equation 53 (see page 25) derives from calculations modifying Gauss' approximation to prime numbers (see Equation 49). Refining the Gaussian approximation utilizes analytic geometry and mathematical logic. The Gaussian approximation tends towards prime numbers over the interval specified in Equation 49.

$$\left\{\int_0^n \frac{dx}{\log(x)} \cong \pi(n)\right\} \tag{49}$$

By using the Gaussian approximation in conjunction with analytic geometry, an algorithm to approximate trigonometric relationships to twin-primes exists. The result is calculable by framing the problem as a distance function between each prime number occurrence, such that Equation 49 by a distance function $(f(d))$ equates to a new value of $[T(\pi(n))]$. From this, Equation 50 results from a conditional after applying the function to Equation 49 (see Equations 50.4-50.8, page 25).

$$\left\{\int \left(\frac{dT(x)}{\log(x)}\right) dx\right\} \tag{50}$$

A definition using equivalence of the distance function to the set {2,2,4} is from a modified number line, which converts the number line into a lattice of three tiers. By establishing a further equivalence between the set {1,3,5} to $[T(\pi(n))]$, though the number (1) is not prime, the distance function is equal to this equivalence. The following Equations 50.1-50.3 show how the algorithm computes the values necessary.

$$[1 + f(2) = 3] \tag{50.1}$$

$$[3 + f(2) = 5] \tag{50.2}$$

$$[1 + f(4) = 5] \tag{50.3}$$

The ordered relationship that may exist works as an assumption to use the anti-derivative of a rule to show that Equation 50 (see page 24) is equal to Equation 53. If a prime number is equidistant from two primes then the integral of a triplet prime may be the area of an acute triangle using the distance function as a calculation for the sides of the triangle. The solution set of the distance function is the twin prime, plus or minus a distance of two, synonymous to a prime number within the finite field specified.

The sum of the distance between a triplet prime and the final distance solution of the set is four. The sum of the distance between the first and last prime in a triplet prime are equivalent within the lattice set for numbers. By expanding the number line into a lattice structure, the base becomes the distance between the first and last prime number. The following premises illustrate this (see Equation 50.4-50.8).

$$[\pi(n)_3 - \pi(n)_1 = d(4)] \tag{50.4}$$

$$\left[\Delta A = \left(\frac{1}{2(4(2\sin\theta))}\right)\right] \tag{50.5}$$

$$[\Delta T = (4\sin\theta)] \tag{50.6}$$

$$\{[T(\pi n) \to [\pi(n) \uparrow f(d)]] = 4(\sin\theta)\} \tag{50.7}$$

$$\int_{\pi n}^{\pi n} \left(\frac{dx}{\log(x)} \pm 2\right) = (0.5 f(d(\sin\theta))) \tag{50.8}$$

From these premises, the following result occurs as Equation 51. Equation 51 and Equation 52 are equal, and simplifies to Equation 53 in complex form.

$$\left\{\lim_{n \pm \infty} n \int_{-\infty}^{\infty} [f_n - f dP_n]\right\} \tag{51}$$

$$\int_{\pi n}^{\pi n} \frac{dx}{\log(x)} + f\{2,2,4\} = \left[\frac{1}{2}(4(2\sin\theta))\right] \tag{52}$$

$$\{ie^{-i\theta} - ie^{i\theta}\} \tag{53}$$

The trapdoor function of the cipher is the result of a function applied to Equation 53 (see page 25) and an integration of entropy. The Pochman expression of Equation 53 (see Equation 53.1) produces a unique analytic function ($\varphi$) (see Equation 53.2, page 27). Figure 1 displays the graph for Equation 53, demonstrating the sine wave with intersections($\pi$). The application of Figure 1 to the system as is engineered allows for use of twin-prime functions along an arc cipher.

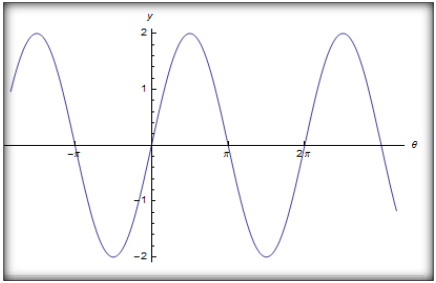

*Figure 1.* ($2\pi$) Period - Function as Viewed in Wolfram Mathematica.

The value of periodicity of the identity shown in Figure 1 from Equation 53 is periodic in ($\theta$) with period($2\pi$). Upon the Pochman expression by virtue of the unique analytic function, when $\{\mathbb{R} \subseteq \mathbb{Z}\}$ and $f(z)$ is equal to ($2\theta^2$) it is calculable from Equation 53. The existence of a single complex analytic function is demonstrable and shown (see Appendix B, page 5).

Equation 53.1 is the Pochman expression of the foundational transcendental identity resulting from modifications to the Gaussian prime approximation. The Pochman expression serves to calculate the existence of the unique analytic function expressed as Equation 53.2 (see page 27). This unique analytic function is critical to proving the removability of the singularity for the trapdoor cipher.

$$\left[ 4\theta \sum_{K=0}^{\infty} \frac{(-1)^K \left( (2K\pi+x)\left(\left(\frac{x}{\pi}\right)_K\right)^3 \right)}{(K!)^3} \right] \tag{53.1}$$

Equation 53.2 is the unique analytic function of which the ceiling value is a Riemannian Manifold. This serves as both an arc function for cryptography as well as the point of radial propagation for the necessary wave functions. Equation 53.2 as the unique analytic function solves for the removable singularity in this system.

$$\left[4\theta \sum_{K=0}^{\infty} \frac{(-1)^K \left((2K\pi+z)\left(\left(\frac{z}{\pi}\right)_K\right)^3\right)}{(K!)^3}\right] = (\varphi) \tag{53.2}$$

Equation 54 satisfies conditions for a *p-adic* vector space and is Figure 2. The combination of using an Abelian-Banach space with ($\mathfrak{H}$) is that by definition these are topological vector spaces, where a vector space is a combination of vectors, satisfying the requirements for Hilbert space. Equation 54 produces Figure 2, which is the initial surface condition of the system cipher and contains entropy in the form of unpredictability.

$$\{2i\theta e^{-ix} - 2i\theta e^{ix}\} \tag{54}$$

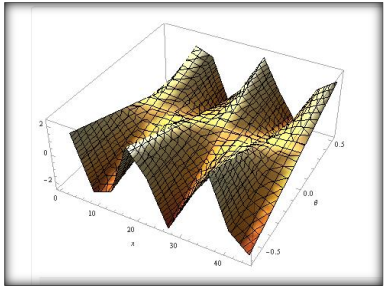

*Figure 2.* ($\boldsymbol{\theta}$) Entropy - Function as Viewed in Wolfram Mathematica

Figure 2 as shown and graphed using Mathematica software using Equation 54 demonstrates entropy in the form of inconsistency of transformations. This serves, with the periodicity of ($\pi$) to approximate the location of prime numbers for cryptographic purposes.

Equation 57 results from a set of parent functions using Equations 55 and 56 along with a variable of ($\pi n$) which relies upon a domain of $\{\mathbb{R} \subseteq \mathbb{Z}\}$ when $\{(\theta = 0), (x = \pi n), (n \in \mathbb{Z})\}$. By applying the function shown in Equation 55, and expanding it into Equation 56, Equation 57 results. Equation 56 once evaluated is undefined where $(\pi n)$, the number of primes, is less than $(n)$ (see Equation 57, page 28). As required for any principle ideal ring, to satisfy this,

Equations 55-57 are applicable to create an ideal ring of which contains an arc for trapdoor functions in the cipher.

$$f: \begin{cases} ie^{-i\theta} - ie^{i\theta} \\ \pi n \\ 2i\theta e^{-ix} - 2i\theta e^{ix} \end{cases} \tag{55}$$

$$f: \begin{cases} 1 - 1 \\ 2\pi \\ 2i\theta e^{-i5} - 2i\theta e^{i5} \end{cases} \xrightarrow{f} \begin{cases} 0 \\ 2\pi \\ ((2e)^{-5i}((i\theta)^{-5i}) - ((2e)^{-5i}((i\theta)^{5i})) \end{cases} \tag{56}$$

$$\{0^{-i\pi n} - 0^{i\pi n}\} \tag{57}$$

By the first set of twin primes $\{2,5\}$ and adjusting Equation 54 (see page 27), the function of Equation 55 accordingly by placing Equation 57 as a member of the function set as $(1 - 1)$, the resulting function is then Equation 58 (see page 37). The algorithmic identity to manipulate the convergence of the complex conjugates and their transposition upon the intersection of a point in the vector space may control the coherence between ultraviolet radiation and radio waves.

Ultraviolet light, explained by the website run by the Nobel Prize committee, explains how computer chips can use the destructive radiation of ultraviolet light in the construction of computer chips. The Nobel Prize organization site reports, "The silicon wafer is moved in steps under the mask and the UV-light to expose the wafer. In this way, chip after chip can be made using the same mask each time" (Nobel Prize Organization, *"Chip Production Today – In Short"* 2003). This suggests the use of ultraviolet radiation to affect the transference of identity based upon discrete functional mapping onto a collinear space for computer electronics.

The support this suggests for the ability of a real-quantity to produce a complex effect from the generation of a bound wave show a capability for the real value to affect the complex conjugate within this system using vector transformation. By the very nature of quantum mechanics, quantum physics uses degrees of probability. The number $(ie^i)$ is a transcendental number where the polar coordinates are $(r = 1)$ and $(\theta)$, the value of polar coordinate degrees. This complex number is a vector in $(\mathfrak{H})$. The use of $(\theta)$, with it serving as the radial center of a pole has potential application to radio antennae using a parametric structure as done in hacking RSA GNuPG keys (Genkin, Shamir, & Tromer, 2013). As conjugate values of $(\lambda)$, the intersection would be between radio waves and ultraviolet light.

**Cyber Security Expert System**

The SWI-Prolog documentation, accessible online, describes the interactive development environment by saying "SWI-Prolog is widely considered to be a robust and scalable implementation of the Prolog language. It is widely used in education and research. In addition, it is in use for "'24 $\times$ 7' mission critical commercial server processes" (SWI-Prolog, 2015). The sets of code for use in this research use artificial intelligence, but development for this research in both cases of the scripts are incomplete. The scripts making use of the knowledge bases and inference engine functions operate according to Ivan Bratko's research, who directed an institute focusing on artificial intelligence (2013, p. 387). The code also uses foundations of Bratko's "best first search" script (2013, pp. 264-268). The knowledge base focuses on ports specific to limited network protocols for the purposes of this proof of concept.

fact:device(input).

fact:device(udp).

fact:device(syn).

fact:device(ipa).

fact:device(port).

fact:(connected(input,port)):-

fact:(connected(port(2),computer2)).

fact:(connected(port(3),computer)):-

fact:(connected(port(4),computer)).

parse:connected(syn,udp,ipa):-parse:connected(syn,udp,syn),input(syn,udp,ipa).

parse:device(syn,udp,ipa).

parse:device(defines,classification,port).

parse:(output(classification(syn|X,udp|Y,ipa|Z))):-input(unknown(X,Y,Z)).

  The purpose of this script is a demonstration of a knowledge base that utilizes an inference engine, which is a form of artificial intelligence. To implement this script in the deployed program for this research requires correctly calling the rules. The gathered arguments for the port scanning which uses the best first search is a recommendation for use. The best first search script has a modification to enable network monitoring upon completion, and therefore incorporates predicates for port scanning. The principle behind the best first search is that the search algorithms do not act traditionally but instead use approximations for the solution to allow faster calculations based on probability for the goodness of fit values (Bratko, 2013, p. 268).

bagof(syn/ipa).

goal(_):-goal(n).

bestf(Vuln,Solution):-

  expand(Vuln,l(Vuln,0/0),9999,_,yes,Solution).

bestf([T|_],F):-

  f(T,F).

bestf([],9999).

expand(P,l(N,_),_,_,yes,[N|P]):-goal(N).

```
expand(P,Tree,Bound,Tree1,Solved,Solution):-port(P),port(Tree|Bound|Tree1;Solved|Solution).

expand(P,l(N,_),_,_,yes,[N|P]):-goal(N).

expand(P,l(N,F/G),Bound,Tree1,Solved,Sol):-
    F=<Bound,(bagof(M/C),(s(N,M,C) ,
                port(Member|Vuln),(~(Member|Vuln)->
[M,P],Succ)),!,succlist(G,Succ,Ts),bestf(Ts,Fl),
        expand(P,t(N,Fl/G,Ts),Bound,Tree1,Solved,Sol);Solved=0).

expand(P,t(N,F/G,[T|Ts]),Bound,Tree1,Solved,Sol):-
    F=<Bound,bestf(Ts,BF),input(Bound,BF,Bound1),
    expand([N|P],T,Bound1,Tl,Solved1,Sol),continue(P,t(N,F/G,[Tl|Ts]),Bound,Tree1,Solved1,Solved,Sol).

expand(_,t(_,_,[]),_,_,never,_):-!.

expand(_,Tree,Bound,Tree,no,_):-f(Tree,F),F>Bound.
```

The rules of the system are the overarching knowledge structure for the algorithmic implementation of this research, which is progress towards programming an expert system for cyber security (Bratko, 2013, p. 347). An expert system must possess knowledge of some form, but also be able to use rules to explain the programmatic behavior to an end user (Bratko, 2013, p. 348). The rule base for the algorithms created operates with a use of dynamic data exchange for Windows operating systems, which allows for communication between applications (Microsoft Corporation, 2015). The dynamic data exchange (DDE) feature within these computations, in combination to the rules shown using best first search actively seeks vulnerabilities through port scans and communication protocols.

```
[trace] 4 ?- '$dde_request'(X,Y,Z,A).
```

X = syn, Y = port(_G2919), Z = ipa(_G2919), A = udp

The principle factor of the algorithms hinge upon the ability to monitor network communication based upon port access and network protocols. This comes from the use of "port" as a predication of several variables.

port(_) :-

    strip_module(port((Module)--> Plain),Module,Plain),

    Plain =.. [Vuln|Args],

    gather_args(Args, Values),

    Goal =.. [Vuln|Values],

    Module:Goal,

    port(port->close).

port(close):-(rl_write_history(port)).

port(classification(on_signal(Vuln|Scan,Vuln|Open,Open))):-(parse:output(Scan)).

port(retractall(Vuln)):-port(Vuln).

port(retractall(parse:parse(Vuln))):-port(Vuln).

port(Open|Scan):-('$dde_execute'((port(_)),Scan,Open)).

((port(Access;Open)):-('$dde_request'((((Access)),write([vulnerabilities]),(Open),(port(_)))))).

(((port(IP)) :-

    dde_current_connection((Scan|Vuln),Scan, Vuln),IP)).

port((_,_)):-'$dde_disconnect'((_,_,_,_)).

Once a current connection opens, further scanning occurs while the port becomes active once again. When a vulnerability registers, further threat analysis occurs. A list of vulnerabilities develops for human analysis and further response. Either the process repeats or the DDE connection can end by user choice. The trapdoor feature of this defense system relies on prime numbers and the DDE capability. For quantum computers, the ability to determine the prime numbers of a factored value renders conventional cryptography pointless. To circumvent this, or attempt to, an infinite recursion based on prime numbers is within the program.

matrix(node(A,B,C),edge([_]),bestf([],9999)):-matrix((node(A,B,C;d(_))),port(A),input(A)).

matrix(Line,Node,Distance):-edge(Line|Node+Distance).

matrix(A|Node_x;(B|Node1,(C|Node3)):-edge(A|Node1),edge(B|Node3), edge(C|Node_x)).

node(d([prime+1=prime])).

node(d([prime+2=prime])).

node(d([prime+1=prime])).

edge(X,Y):-(matrix(lattice,([])|X,Y)).

edge([Node1,Node2];[(C;Node3)],[_]):-matrix(Node1|_,Node2|C,Node3).

edge([A,B];[B,C];[C,B]):-node(3),edge([A,B,C]),distance((node + edge =Distance)),matrix(edge,node,Distance).

The principle behind the infinitely recursive prime lattice structure is a distance function between prime number locations on a natural number line, but upon a modified number line. The traditional natural number line is a single line where each natural number has an equidistant position, but the lattice matrix of natural numbers developed for this research is in place of that.

**Virtual Quantum Circuits**

The phase shifts for building this quantum circuit are several inversion functions. The inverse Quantum Fourier Transform (QFT) is upon $(x)$ during $|0\rangle$, while an inversion carry function is at $|1\rangle$ upon $(x)$ leading to transfer of spin through the vector space. Simultaneously an inversion reverse function within $|1\rangle$ upon $(y)$ is in place. Finally, a swap inversion function is at

|1⟩ upon (z) while a control gate during |0⟩ upon (z) results in final superposition. Figure 3 displays the initial conditions and state space of the quantum circuit.

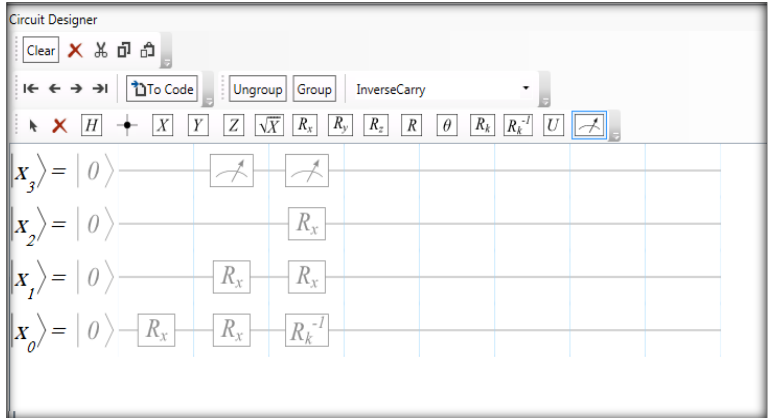

*Figure 3.* Quantum Circuit - Circuit as Viewed in the QuIDE Environment.

Figure 3 depicts the initial qubits in the QuIDE after adding the described inversion gates and measurements. The result after adding these specific gates was intended to create a chain effect of transferring states such that a model of communication results which rely on entanglement and coherence. Figure 4 is the ordered result of Figure 3 after selecting the "build circuit" option in QuIDE. This demonstrates the fidelity of gates selected to operate as a quantum circuit.

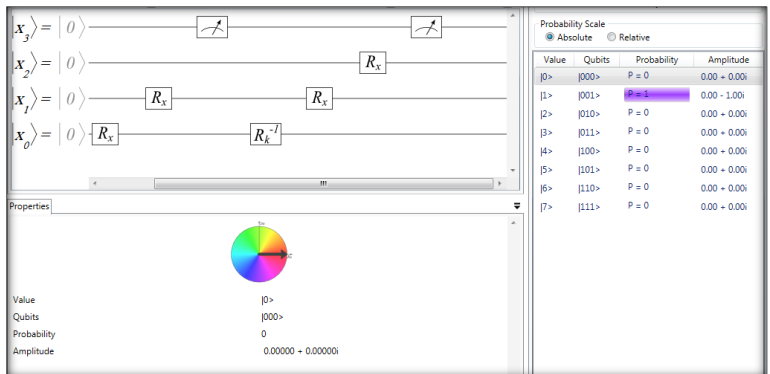

*Figure 4.* Phase Shifts – Circuit as Viewed in the QuIDE Environment.

From the configuration shown in Figure 3 as built, Figure 4 shows how the QuIDE program sorts the gates along with the measurement options once the circuit compiles. The following code populates when selecting the "build circuit" option. The rotations shown in the "x.RotateX" pieces of the code are rotations upon the *x-axis* with a degree of ($\pi$) values. The measurement functions are relative to the inverse phase shifts and occur both before the shift and after.

```
using Quantum;
using Quantum.Operations;
using System;
using System.Numerics;
using System.Collections.Generic;
namespace QuantumConsole
{public class QuantumTest
{public static void Main()
{
QuantumComputer comp = QuantumComputer.GetInstance();
Register x = comp.NewRegister(0, 4);
x.RotateX(3.14159265358979, 0);
x.RotateX(3.14159265358979, 0);
x.RotateX(3.14159, 1);
x.Measure(3);
x.InverseCPhaseShift(3, 0);
x.RotateX(1.5707963267949, 1);
x.RotateX(1.5708, 2);
x.Measure(3);
}}}
```

## Analysis of Results

After testing, to understand the standard operations that result from the selected quantum gate configuration the initial collapse, where the probability of the qubit value is $(1)$, followed by the inversion upon that qubit value, is a function of probability mapping onto another qubit. This is proof of fidelity to the research discussions in the previous sections. The results are a combination of mathematics, physics, output from the algorithms engineered, and output based upon the virtual quantum circuit built for this research. The most impactful result is a combination of the capabilities of these computations (see Appendices A and B) along with the implications for mitigation against a perceived threat that quantum computers may pose.

**Complex Convergence and Polar Coordinates**

A singularity at the intersection of $(y)$ and $(\theta)$ between the complex conjugates and real components approach values supporting the findings of Turchette and their team's conclusions of quantum computing scalability (Turchette, et. al, 1995, p. 4711). Once the near-value supporting Turchette and their team's findings became discoverable, the results of calculation for this algorithm's functions became applicable for quantum computing implementation in addition to classical system defense. While an attempt may be to use classical programmatic defense, given the potential for a quantum computer to easily circumvent these conventional cyber security measures, additional mitigation is proposed in the form of hardware using the principles of radial wave mechanics as the triggered execution of a trapdoor function. The findings from setting the identity to $(n)$ when $(n)$ vary between the ranges $[-5, ..., 5]$ with the variable at $(x = 5)$, produces Figure 5 and Equation 58 as a generalized formulation of the initial system state.

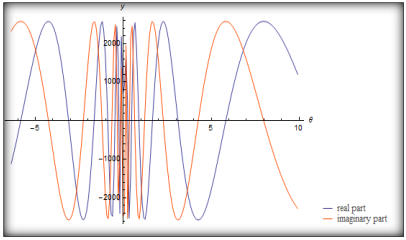

*Figure 5.* Singularity in $(\boldsymbol{\theta}, \boldsymbol{y})$ - Function as Viewed in Wolfram Mathematica.

Figure 5 demonstrates the ability for a complex function to result in a convergent series, which supports the findings shown by calculation of Equation 58.

$$\left\{\left(-n^{-i\pi n}(-1 + n^{i\pi n})(1 + n^{i\pi n})\right) \in (\lambda y = fn)\right\} \tag{58}$$

The difference along $(\theta)$ comparative between the initial entropy of this system and the discoverable complex singularity is removable and attenuated for. Analyzing the wavelength of the *y-axis* as graphed in Figure 6 shows a convergence of the real and imaginary components at the values of $[(-1.45 * 10^8), (2.0 * 10^5)]$ which accordingly has a vector length of approximately $(-1.45 * 10^8)$, a horizontal angle of $(179.21°)$ and vertical angle of $(89.21°)$. The value of $(\theta)$ thus equals $(179.21°)$ as shown by Figure 6.

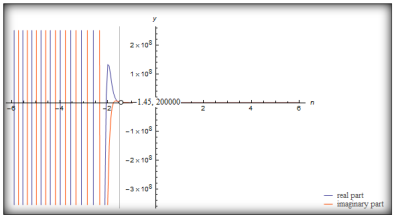

*Figure 6.* Complex Intersection – Function as Viewed in Wolfram Mathematica

The node at the closure of the wave in Figure 6 has an amplitude equal to a resulting rate of coherence from Turchette and fellow researchers in their findings (Turchette, et al., 1995, p. 4711). Applying the degrees of $(\theta)$ as a value in the polar coordinates of Figure 7 expressed as Equation 59 creates a radial propagation mechanic.

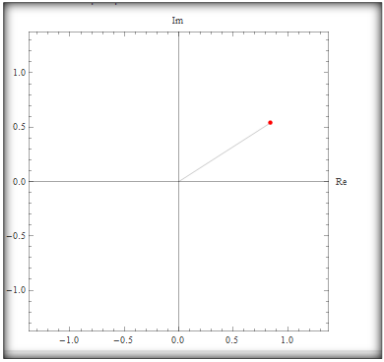

*Figure 7.* Polar Coordinates – Function as Viewed in Wolfram Mathematica

$$\{ie^{179.21°(-i)} - ie^{179.21°i}\} \tag{59}$$

The number $(ie^i)$ is a transcendental number where the polar coordinates are $(r = 1)$ and Equation 60 is the polar coordinate value of $(\theta)$ in degrees. This complex number is a vector in $(\mathfrak{H})$. Equation 60 is the value of substitution for $(\theta)$ in this system and is the center of propagation for engineering radial network defense.

$$\theta = \left[\frac{180\left(\frac{\pi}{2}-1\right)}{\pi}°\right] \tag{60}$$

The use of Equation 60, serving as the radial center of a pole has potential application to radio antennae. When $(\theta)$ is set at $(180°)$ in the exponential value for degrees, the results are an intersection approximate to a zero-value equaling $(2.0 * 10^{-2})$ and $(2.45 * 10^{-16})$ as values of $(\lambda)$, thus the intersection would be between radio waves and ultraviolet light. The use-case under discussion requires the coherence between such waves, which in turn requires the satisfiability of removing the singularity at the intersection of $(\theta, y)$ upon the *y-axis*.

**Automated Cyclic Port Forensics**

Figure 8 illustrates how the cycling of port scanning operates from the computations (see Appendix A). By equating the variable of "Vuln" for "vulnerable" to the argument or "Args" of [_S1], a request to the user in identifying whether the port specified is vulnerable becomes a capability of *QUINE*. With no human input, the next step in this process is to begin the best first search.

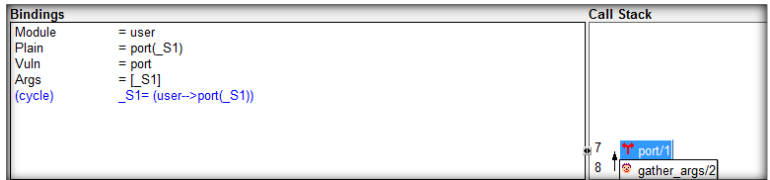

*Figure 8.* *QUINE* Port Bindings – Output as Viewed in the SWI-Prolog Debugger

The ability of the algorithms to isolate useful forensic data along with data relevant to network defense is shown by output which comes from entering the request of "gather_args(X,Y)." in the prolog terminal. The option to trace calls of predicates and variables starts when the user enters "trace." into the terminal. The following is the listed output from a trace that comes from the gather_args query, which results from the algorithms of *QUINE* (Malloy, 2015):

X = Y, Y = [] ;

  Redo: (7) gather_args(_G8151722, _G8151723) ? Listinggather_args([], []).

gather_args([+A|C], [B|D]) :- !,

    unknown(port(A, B)),

    gather_args(C, D).

gather_args([A|B], [A|C]) :-

    gather_args(B, C).

gather_args(port(A), port(B)) :-

    on_signal(A, B, _),

    port(A),

    port((B| A)).

gather_args(file(D, E), G) :-

'$append'(A, [tuple('All files', *.*)], B),

A=..[chain|B],

current_prolog_flag(hwnd, F),

working_directory(C, C),

call(get(@display,

    win_file_name(D,

        A,

        E,

        directory:=C,

        owner:=F),

    G)).

win_menu:gather_args([], []).

win_menu:gather_args([+A|C], [B|D]) :- !,

    gather_arg(A, B),

    gather_args(C, D).

win_menu:gather_args([A|B], [A|C]) :-

    gather_args(B, C).

    Figure 9 designates the port scanning as the "Best-Port First Search" (BPS) given that the reference of "best" is to the optimized search function along with the principle of an attacker desiring the most vulnerable entry point.

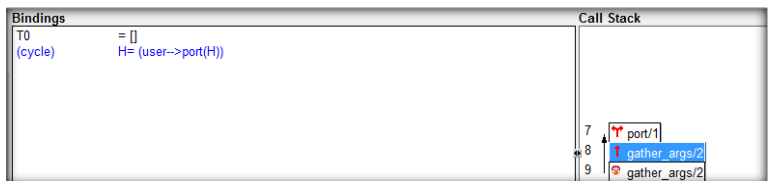

*Figure 9.* QUINE Port Arguments – Output as Viewed in the SWI-Prolog Debugger

The binding of T0 to [] is a signifier of both the ability to again structure lists from the scan as well as the capability to further refine the use of the best first search. This is possible by restructuring the computations dynamically at runtime if one chooses to do so. If required, the ability to set T0 to X from the output listed by the gather_args query, in addition to the [_S1] argument shown in Figure 8 (see page 39) as a binding to a port, the hypothesis of the list resulting is of forensic information associated with that port.

**Virtual Coherence and Entanglement**

The C-NOT operation from the selected quantum gates for this circuit adds to systemic validity of this quantum circuit model. To retain fidelity to the research Turchette and the researchers performed, developments require coherence and entanglement (Turchette, et al., 1995, p. 4714). Thus, Figure 10 illustrates the fidelity between this research and the requiments set forth by Turchette's team for the measurment of conditional phase shifts in quantum computing (Turchette, et al., 1995, p. 1411). Figure 10 shows coherence between the qubits ⟨000|110⟩ along with the superposition of qubits ⟨010|100⟩.

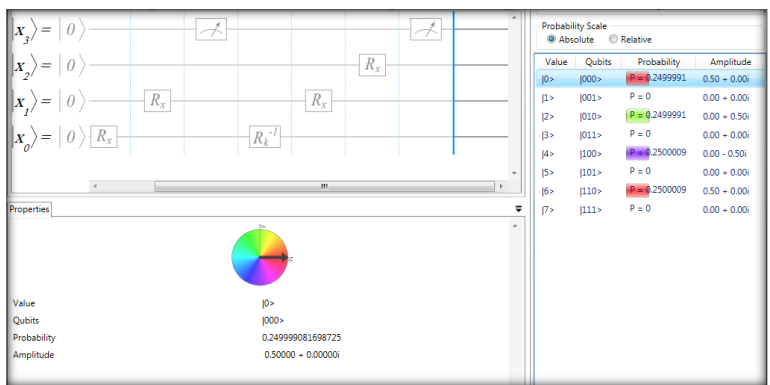

*Figure 10.* Superposition and Coherence – Circuit as Viewed in the QuIDE Environment.

Figure 10 has a probability allowing strong confidence in the amplitude with a value of $(-1.0 * 0.0i)$, while the effect upon it is essentially a form of C-NOT operation given that the qubit in the initial state was |000⟩ yet the collapse in probability is upon a new qubit |010⟩ as

required according to Equations 2-6 (see page 5). The final confirmation that coherence exists within the quantum circuit is supported by the coloring rules of the qubits ⟨000|110⟩ at values ⟨0|6⟩ in addition to the superposition seen from the coloring rules of ⟨2|4⟩ (see Figure 10).

By pursuing the principles behind logic bombs in conjunction with developing necessary system conditions for quantum computing, the results in this research set forth critical aspects to any form of mitigation against quantum computing, or the deployment of security systems based upon quantum computing. The applications of inversion are possible using logical structures within *QUINE*. Ivan Bratko's code, despite the modifications is specifically an application to create a knowledge base and report using inference (2013, p. 386). *QUINE* operates with inversion principles using concatenation functions. This concatenation, though perhaps useful to some degree is not an accurate implementation of quantum computing algorithms. Figure 11 shows the utility in key bindings and inference capable by the algorithms. From Figure 11, Figure 12 results, producing classification of determined variables in network communication from an unknown argument. Figure 11 is the internal reasoning function of *QUINE* as is producible by the algorithms implemented.

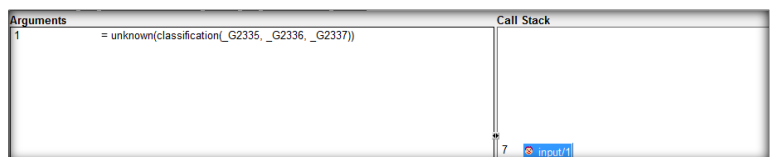

*Figure 11*. Classification Argument – Output as Viewed in the Prolog Debugger

The conjunction of the modified knowledge-inference engine with the mitigation script may further implementing the computations to isolate identifiable malware hashes or perhaps even create new hashes to mitigate polymorphic viruses autonomously. Figure 12 shows the final step of the arguments that follow from the internal reasoning of the *QUINE* expert system knowledge inference engine.

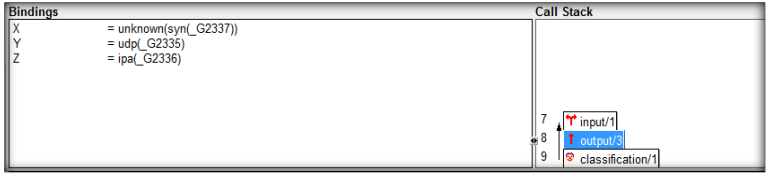
*Figure 12.* Variable Classification – Output as Viewed in the Prolog Debugger

The output of *QUINE* in Figure 12 demonstrates the capability for tactics relying on automated or human involvement. Should this schematic of an expert system enter use, humans must remain in control of each decision, for not only resilience, but safety as well.

**Wave Filters and Encryption**

The conjunction of inherent entropy to the surface throughout the system is a matrix transformation of the applied tensors. This surface tension with a rotation upon the central radial propagation mechanic produces Equation 61. Equation 61 is a period of prime locations upon the *x-axis* with a root of complex identity. The period $(2\pi)$ and root $(n)$ as element of $(\mathbb{Z})$ is periodic in $(x)$ and exhibits promising wave propagation mechanics.

$$\left\{ 2i \frac{180\left(\frac{\pi}{2}-1\right)}{\pi} \circ e^{-ix} - 2i \frac{180\left(\frac{\pi}{2}-1\right)}{\pi} \circ e^{ix} \right\} \tag{61}$$

The graphs shown illustrate that a complex wave function through the *y-axis* may possess two simultaneous wave propagation patterns that are an active interaction with the respective imaginary component. The wave propagation patterns are promising findings from this research. The pertinent resulting algorithmic expression for the construction of a radial-wave shield graphically depicts the interaction of complex wave functions. Further analysis suggests a consistency of the set of identities resulting from this research to some parent function of transcendental and complex identity. Equation 62 is an additional transcendent identity for algorithmic implementation as well as radial wave shielding with engineered antennae.

$$\left\{ e^{-ix} - 2i \right\} \tag{62}$$

Figure 13 implies an ability of a relative interaction between the real values and complex conjugates. The relationship between interactions of wave mechanics as shown by Figure 13 illustrates the corresponding elements of the real and imaginary components to the complex identity of Equation 62 (see page 43). The ability for coherence and entanglement within the wave function of Figure 13 is an algorithm, which produces the surface tension allowing for interception of malicious signals.

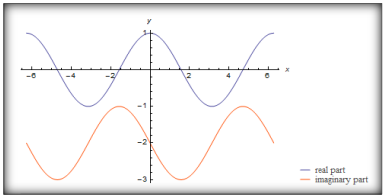

*Figure 13.* {$\Re$ , $\Im$} Correspondence – Function as Viewed in Wolfram Mathematica

The difference in wavelength between Figure 14 and Figure 13 suggests the ability for radial shield mechanic as an algorithmic identity to manipulate the convergence of the wave function of the *y-axis* with the initial entropy of the cipher inside the principle ideal ring. Figure 14 exhibits excitation of phase-states within the higher-order tensors.

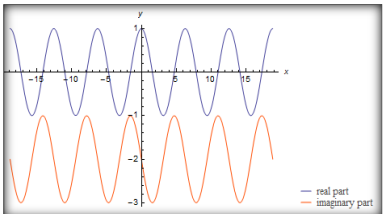

*Figure 14.* {$\Re$ , $\Im$} Mapping - Function as Viewed in Wolfram Mathematica

The resulting support suggests the ability of a real-quantity to produce a complex effect from the generation of a bound imaginary wave and shows a promising capability of wave mechanics to be implementable as engineered for this research. This forms the coherence between ultraviolet radiation and radio waves graphically displayed by real and imaginary bijective correspondence. This is strongly suggested by Figure 15. Figure 15, as shown by the rotatation displayed, implies a wall effect as a general surface from a perturbation of $(n)$ values along $(x)$. Figure 15 of Equation 63 is the effect of a curl in the vector $\vec{(j)}$.

$$(n^{-i\pi x} - n^{i\pi x}) \tag{63}$$

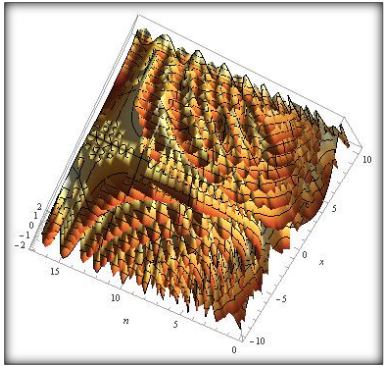

*Figure 15.* Curl Decomposition – Function as Viewed in Wolfram Mathematica

Figure 16 results from Equation 63 and for some $(n \in x)$ with a rotation $(\theta)$ a manifold thus allows elliptic curve cryptography (ECC) from this system. In terms of ECC, there is more necessary. Figure 16 is curling of the *x-axis* which is the vector of light propagation expressed by a transcedent algorithm which adjusts the theta values upon the imaginary number $(i)$ such that the theta value at $(0.1)$ exhibits characteristics of decomposition.

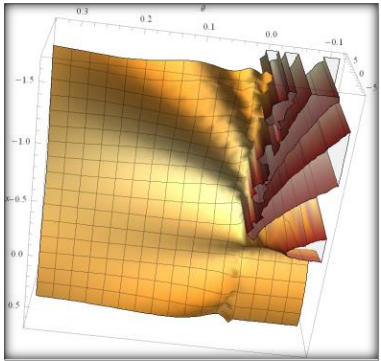

*Figure 16.* $(\nabla x_j)$ – Function as Viewed in Wolfram Mathematica

Figure 17 shows an elliptic curve near the center of the graph as a rotation of $(x)$ along the center of the manifold. Equation 64 is the generator function of Figure 17. ECC operates using a generator function such as this; however, Equation 64 has a cofactor of at least two.

$$(2e^{-(i(\pi x))} - \theta^{i(\pi x)}) \tag{64}$$

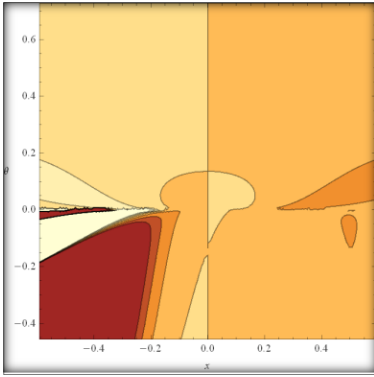

*Figure 17*. ECC Curl – Function as Viewed in Wolfram Mathematica

Figure 17 demonstrates where upon the field a complex number may lay, the least lower bound (LLB) is a complex root of this system. The LLB is a curl of lambda upon the *x-axis*, which is a complex vector $(\vec{j})$ across the vector space. The curl of the vector is a function, which shows a curl($\nabla x_j$). The vector $(\vec{j})$ at $\sim(-0.2,0)$ creates a point of propagation to mitigate incoming malicious signals. Figure 18 illustrates the wavelength($\lambda y$) in coherence with the ultraviolet light spectrum.

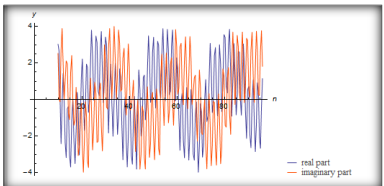

*Figure 18.* ($\lambda y$) – Function as Viewed in Wolfram Mathematica

The plot of ($\lambda y$) shown by Figure 18 (see page 46) results from Equation 65. The propagation mechanic exhibits the behavior of the creation of light given the vector upon which the trajectory follows.

$$\{2e^{-i\pi n} - 2e^{i\pi(n)}\} \tag{65}$$

The trajectory of ($\lambda y$) is a curve between the interval $(\alpha, \beta)$ where the path towards the point($\alpha$) is a vector of light generation. The point($\beta$) is the bottom eigenstate of this system and is expressed by the radial propagation mechanic such that Equation 66 results.

$$\left\{-\frac{ie180°(\pi)}{\sqrt{2}} + e^{-180°i}x\right\} \tag{66}$$

Where $(n = E)$, and $(0 < y < 1)$ Equation 67 is the resulting propagation and exhibits excitation of the wave function. Figure 19 illustrates the phase shift within the wave function.

$$\{(e^{-ix} - 2i)\} \tag{67}$$

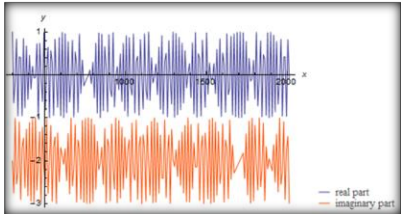

*Figure 19.* $(\lambda x)$ – Function as Viewed in Wolfram Mathematica

The algorithmic removal of the singularity within this system serves both as a function of trapdoor capabilities within the ECC cipher as well as the handshake method. This expression is the attribute of the zero of the system, which is analogous to a NULL, or blank value within the RSA 4096 public key cipher. The removal of the singularity is a demonstration of further feasibility in the application of the *QUINE* algorithms in conjunction with the mathematical identities so engineered. The solution to remove the singularity is Equation 68.

$$\{2e^{-i\pi n} - 2e^{i\pi(n)}\}, \tag{68}$$

Figure 20 is the point of intersection at the zero of the system that demonstrates a node of coherence. This is a graphical representation of the solution to remove the singularity by virtue of Equation 68 (see page 47) along the *y-axis* as a variable of $(n)$ value. Figure 20 is a wave function of a value $(n)$ upon the *z-axis*.

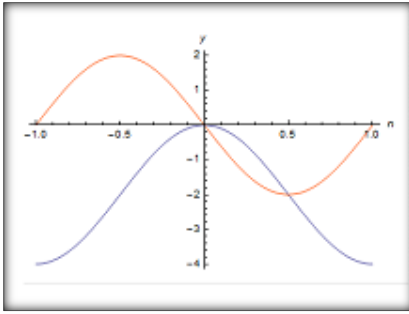

*Figure 20.* (λz) – Function as Viewed in Wolfram Mathematica

The following integration of Equation 69 over the interval $(-1,1)$ results in a Taylor expansion at $(x=0)$ in$(z)$.

$$\int_{-1}^{1}\left\{\begin{array}{c}\left(\left(2\exp\left(-(i(\pi x))\right)-\frac{\theta^{(i(\pi x))}}{(2i\theta e)^{-ix}}-(2i\theta e)^{ix}\right)dz\right\}\\=\\\{2(-(2e)^{ix}\theta^{(i(\pi x))}(i\theta)^{ix}-(2e)^{ix}(i\theta)^{ix}\\+(2e)^{i\pi x}\}\end{array}\right\} \quad (69)$$

The final resulting capabilties of this system are demonstrated in terms of a Riemann-Hilbert intersection, where the creation and subsequent propagation follows upon a curve such that the resuling field is homomorphic and surrounds specific coordinates upon the *x-axis*. This is represented by Figure 21, which is a mapping of Equation 70 (see page 49).

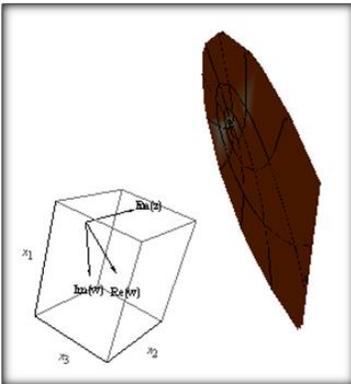

*Figure 21.* Riemann-Hilbert Intersection 1 - Function as Viewed in Wolfram Mathematica

The ceiling function of the complex conjugate upon the cube of pi, as subtracted from the calculated existence of a unique analytic function (see Appendix B, page 5) produces the necessary gravitational analog serving as the trapdoor of the cipher. This is expressed as Equation 70. The shield propagation as a surface tensor through the complex identity of Equation 70 upon the *x-axis* is significant proof-of-concept for a radial, standing wave mechanic.

$$\{\lceil \varphi \rceil = \left(12.511 - \left(\frac{z^4}{\pi^3}\right)\right)\} \tag{70}$$

Equation 71 is a proportional integration of the vector space in conjunction with the final, unique complex analytic function. The demonstrability of the use-case for Equation 71 is expressible by Figures 22 and 23. Figure 22 demonstrates the complex roots of the plane, as well as the transformational capabilities of the manifold. Equation 71, which produces Figure 22, is representative of the boundedly compact manifold necessary for subsequent mitigation.

$$\left\{\left\{-\frac{z^4 - 387.9}{\pi^3}\right\} \propto \left\{\frac{d}{dz}\left(4(179.21°) - \frac{z^4}{\pi^3}\right) = -\frac{4z^3}{\pi^3}\right\}\right\} \tag{71}$$

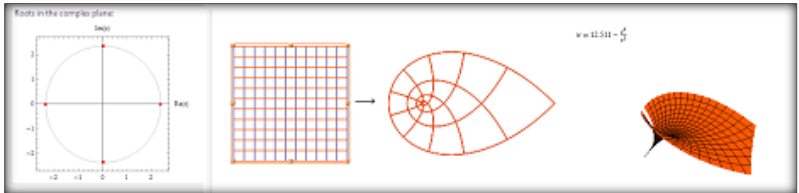

*Figure 22.* Riemann-Hilbert Intersection 2 – Function as Viewed in Wolfram Mathematica.

Figure 23, which is producible by the same unique analytic function, is a Riemann sphere mapping that denotes a particle and wave duality with collapse. The collapse of this function must serve as the triggered propagation pattern of coherence between the complex algorithms discussed in this research. The sympletic space produced within the satisfied (ℌ) as engineered, conclude the proof of concept in the feasibility of implanting the radial dynamics by virtue of higher-order tensor products of a principal ideal ring. Figure 23 denotes the producible shield within an orbital surface structure.

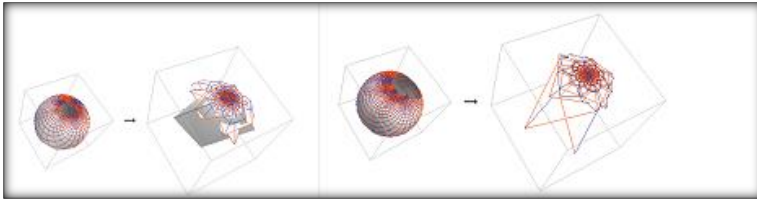
*Figure 23.* Riemann-Hilbert Intersection 3 – Function as Viewed in Wolfram Mathematica.

Figure 24 is a further transformation of Figure 23 in that the mesh of the corresponding values upon the *z-axis* is reducible to a smoother structure. This demonstration is in accordance with the principles of Gaussian channels. The bandwidth allows manipulation such that transmission with reduced error or "noise" creates a dual-purpose channel within the same spectrum.

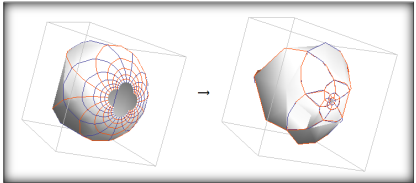
*Figure 24.* Riemann Sphere Map – Function as Viewed in Wolfram Mathematica.

Figure 24 as depicted is the action principle within the removable singularity solved for in this work. The group topology of the hyperdimension in Figure 24 signifies the necessary boundary conditions for radial wave mitigation against acoustic threats. By virtue of this research, acoustic mitigation via methods of coherence in radial wave mechanics is feasible.

## Discussion of the Findings

Networks undergo constant threat, and evolving mitigations are a necessity. The difference in wavelength calculated within the structure of the novel system developed suggests the ability for the wave functions of light as radio frequencies tending towards the ultraviolet spectrum to interact predictably within the confines of this system. The polar coordinate and wave mechanics embedded as a trigger event must create a wave propagation. The use of the trapdoor function as a trigger event to begin wave propagation would be a mitigation against the classical acoustic attacks to break RSA encryption.

The combined event of radio wave propagation along with the series expansion within the trapdoor is a full implementation of this defense system. The infinitely recursive lattice within *QUINE* activated upon a prime number value can be coupled with the node at which a key value prime is factored by an attack. To implement the system as a successful defense against quantum threats should not be retroactive or reactive in use.

The threat of a quantum computer attack against a conventional system poses a risk of degrading traditional communication and operations of national critical infrastructure if the quantum computer should be under malicious control. Shor's algorithm is a benchmark for the performance of a quantum computer on a large scale. For cyber security mitigation in quantum-computing networks, the benchmark of mitigation should be predictive analytics, and the scalability of such must be a factor of resilience.

**Classical Defense against Quantum Threats**

Entangled and coherent attacks propagating from quantum computers may hypothetically possess the ability to enter entanglement with a targeted device maliciously. To construct a defense system to mitigate such potential threats, using a system of propagating radio waves to forcing continuous decoherence around a defended network may shield against attempts of malicious targeting. The applications of the ultraviolet light, as a possible defense component, rely on the electromagnetic aspect of ultraviolet light. Ultraviolet light cannot ionize an atom, but the properties of radio waves, another form of light, can couple with conductors if the distance is within the propagation of the wave. Thus, the exchange of radio waves with ultraviolet light may alter hardware.

Hilbert spaces are requirements to any quantum based mathematics for applications to computation, and therefore must be present in some way for security systems relying on quantum mechanics. During the initial stages of this research, the power of quantum computers understood in terms of threats to cryptography was not the focus, but applications for defending against quantum cracking is worth noting. Users of the Internet may not know that the transmission of data between their computer and the websites they visit rely on strong encryption to protect them, but with quantum computers, encryption using RSA and similar algorithms cannot mitigate cracking by a quantum computer. The promises of advanced elliptic curve cryptography suggest greater potential to mitigate the threat of quantum computers to classical encryption.

Given the fact that classical computers themselves operate using quantum mechanics, but fall to the limitations of classical systems, it is apparent that with the research conducted by Toshiba and their affiliates at Cambridge, quantum computing can affect classical systems. The level of threat posed by this finding, in conjunction with the currently occurring race to construct quantum communication, emanating from satellites, provides support for the reasoning behind use of electromagnetic waves for computer network defense. Given Earth's magnetic field inhibits quantum communication, conceptualizations of energy excitation using radio waves to create a shield appears tenable.

For instance, the acoustic hacking of the RSA GNuPG key suggests a potential adjustment of radio frequency emitted by a CPU to act as mitigation, not a vulnerability. If such mitigation fails by transitioning from a hyperfine state as demonstrated by Turchette and their research, to a vector space of reflection using a curl along the orbital axis of momenta a defender can feasibly cause decoherence of the malicious signal or destroy the malicious signal. Should the mitigation succeed, the defensible attack should incorporate the research done from Cambridge and Toshiba where conventional telecommunication fibers carry quantum information. If the Cambridge-Toshiba experiment is repeatable, the ability to extract a GNuPG RSA key from a target system is feasible with quantum communication using classical devices. The next line of defense is a next-generation seed algorithm for elliptic curve cryptography.

Applications of space as a vehicle for quantum communication may be promising, but is not necessary in all respects. In addition, with cryptography using the principles of elliptic curve geometry, of which can be complex the ability to conjoin the principles behind orbital momenta, a vector space, and trapdoor functions of path integrals over the space can be an advancement of seed algorithms for elliptic curve cryptography. Path integrals in quantum fields over vector spaces may provide an intractable problem applicable to mitigating attempts to break encryption.

The intractability may arise given that the paths themselves are vectors within a vector space. The inability to predict the seed value compounds when using pseudo-random transformations of the area under the curves from the motion of points. The end goal is a defensive system such that any quantum computer attack against a conventional computer would trigger an automatic response using combinations of conventional hardware and quantum based computation. The distribution of a propagating wave as mitigation will suffice to disrupt quantum computer threats using wave-particle interaction.

**Vulnerability of RSA 4096 Key Cryptography**

The vulnerability of the RSA cryptographic system extends further than emission of key values through acoustic leaking, but also within experimental protocols ubiquitous to certain browsers. The vulnerability within RSA is the foundational algorithms itself, which assume AES in certain ubiquitous protocol deployed, yet not secure. Figure 25, for the purposes of this discussion, is an example graph of network communication generated from a comma-separated value. An analysis reveals a pattern of communication between end-user and server.

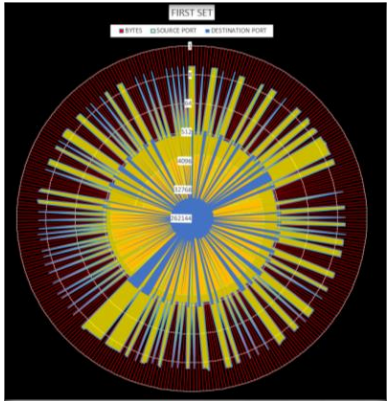

*Figure 25.* RSA Set 1 – Function as Viewed in Microsoft Excel

Figure 25 when analyzed further produces Figure 26, which is a second set of the same network traffic, though reduced to a specific protocol. The pattern of communication becomes more apparent along with values for factorization. Upon reduction of cipher values, an algorithmic methodology generalized to crack RSA ciphers is possible to the degree that the value of the key length as a function of a unique division results in the key system's exchange, length, and values.

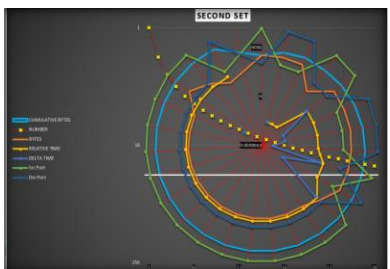

*Figure 26.* RSA Set 2 – Function as Viewed in Microsoft Excel

Figure 26 necessarily leads to a key decryption based upon the factorable values inherent in RSA key exchanges. The resulting Figure 27 illustrates the handshake of the network communication as defined by the protocol under analysis. Figure 27 shows fully permissible with the methods employed how the null values and "blank" spaces are the start of the headers for the cipher.

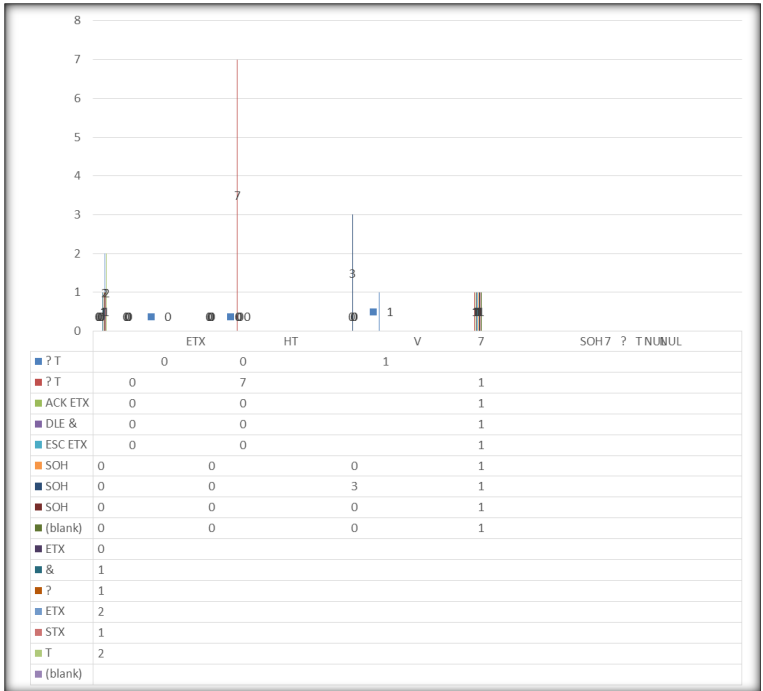

*Figure 27.* RSA Crack 1 – Function as Viewed in Microsoft Excel

Figure 28 shows the ASCII values of the now decrypted RSA 4096 cipher under analysis. The vulnerability as shown exists through the entire communication between end-users of this browser, which utilizes an experimental protocol.

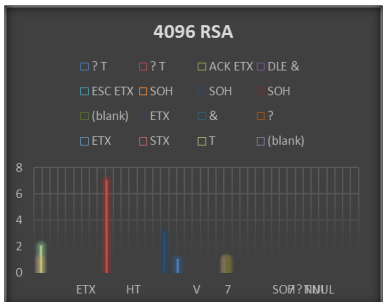

*Figure 28.* RSA Crack 2 – Function as Viewed in Microsoft Excel

With the ASCII values determinable as illustrated by Figure 28, Figure 29 shows the full handshake pattern of this specified RSA cipher. The ability to determine this required very little in methodology, once the comma-separated values are extractable from network traffic generated by non-abnormal browsing such that the only destination was the home page of the browser, which created this protocol.

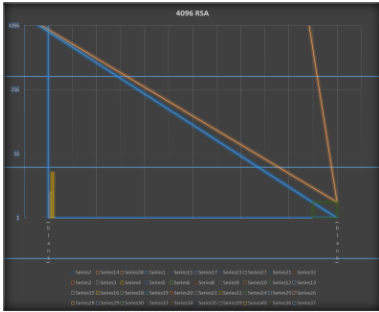

*Figure 29.* RSA Handshake and Key – Function as Viewed in Microsoft Excel

With the entire RSA 4096 cryptography now mapped and cracked, it is evident that the vulnerabilities within this are in need of both reevaluation and removal from operation. The suggested replacement is feasible and is supportable by this body of work, though further refinement and development is necessary.

**Complex Elliptic Curves and Signals**

The NSA understands the threat posed by quantum computers against encryption because the impact it has on sensitive material owned by the United States government. In terms of the threat posed by a quantum computer, there is no limitation in attacks it could perform, or to the computations, a quantum computer can perform. At the heart of computation and at the heart of physics and mathematics are logical relationships and operations. Physics is reducible to mathematics, and so is quantum computing. Therefore the research focused mostly on the equations surrounding the principles of physics and mathematics as well as the conditions required for computational satisfiability. This is to aid in both understanding for, and development of mitigation capabilities for tactics, techniques, and procedures in computer network defense.

The application this research has for policymakers is difficult to address. This research allows classification of what constitutes a quantum computer or not, should they ever be limited to governments and academia for use and operation. Should a cybercriminal ever construct a quantum computer for malicious purposes the effects would be catastrophic for any victim. Instead of having to spend time infecting target machines, the infected machines could possibly suffer an attack instantaneously with entanglement, and transport any information from the infected machine to the attacker instantaneously based on coherence.

In as much as a physical attack against a computer destroys the logical operations of it, logical operations can destroy physical targets. Defense and mitigation measures to protect computers from physical and logical attacks vary in success, but are extremely difficult to maintain without considerations of resilience. Quantum computers of 2015 will most likely remain the research focus of academics but may soon be the tools of governments. Even if this proves accurate, the threat of a malicious criminal managing to command and control a quantum computer supersedes a nation-state threat.

Quantum computer networks for communication do not need traditional network security monitoring tools given the use of coherence and entanglement. By the very nature of quantum communication, there is an immediate alert of any would be eavesdroppers and the data is destroyed. The wave function of the particle given the energy and the momentum dictates aspects of the behavior of the wave function. Any act of observing the communication alters the position, or momentum, which notifies Alice and Bob of an attack. There is potential for mitigation using conventional programming with a Gaussian channel using a synchronous concurrent algorithm so the discrete time interval and vector manipulation is analogous to a logic bomb's trigger event. In place of a malicious result from the logical trigger, an automated call function of the correct argument could hypothetically implement a logical phase shift in a traditional cyber security system. With this manner of defensive mitigation in place, the need for final human authority would be critical to prevent a malfunction of the automatic response.

**Expert Security Systems**

The use of an expert system or even aspects of expert systems for use in cyber security are an implementation of artificial intelligence. The use of artificial intelligence in cyber security may cause concern to some, though the actual implementations, which are possible, do not create a significant threat. The need for human interaction with cyber security comes with the implementation of resilient network defense. This remains true with systems that have artificial intelligence incorporated in the algorithms. The difference between automation and artificial intelligence is the system with artificial intelligence uses reasoning and can explain the reasoning to a human. A firewall uses a rule-based system to act against threats, but an expert system is capable of performing an action that an expert in the field would choose.

The human behavior behind an attack is unpredictable to a large degree, as much as any human action is. The ability for an expert system to conduct any form of network operations requires predictability of human reasoning to some extent. A full expert system conducting network operations would not be a safe implementation of artificial intelligence. An expert system that informs a human in decisions that would supplement network operations is a safe implementation of artificial intelligence. If a fully intelligent and autonomous application were to malfunction, the results could be a perceived mitigation towards false-positive threats both external and internal.

An internal action of mitigation by an artificially intelligent application if malfunctioning as an "implosion" would be the internal system destroyed in a false-positive response. The converse of that, an "explosion" would be an instance of the same network defense system perceiving all outside networks as a threat and attacking those systems. In either situation, a simple control mechanism of human authorization would prevent each form of malfunction by preventing a false-positive trigger.

**Limitations**

The necessary background in understanding the mathematical and physical components of quantum computers, allowing strategic defense to be developed is incomplete. SWI-Prolog is not simple to deploy in all situations. While those with strong backgrounds in formal logic may find writing code in prolog somewhat straightforward, there can be issues such as with this script due to cumbersome variables distributing predications in unexpected ways. For example, the results when attempts to test the DDE request feature targeting a port produces an issue of values equivalent between port and "ipa," which stands for Internet protocol address (IPA). While this may not be a fatal error, given that it may attribute a port under attack to the IPA the attack is emanating from may also erroneously mark the port equivalent with the IPA. As with any research, the greatest limitation is the time allotted for conducting such studies.

**Recommendations**

Future efforts making use of the research conducted within this thesis reach beyond the field of cyber security. First, the effort uncovers a method by which quantum gravity as a union of Einstein's relativity and quantum understandings of wave-particle duality is both predictable and testable. This is possible by methods of acoustic attacks and subsequent triggering of the acoustic shield, thereby drawing the malicious wave into the shield's field within the Riemann-Hilbert intersection. This is analogous to the curvature of space-time relative to a gravitational field.

Secondly, and explicitly applied to security, are explorations of attack and defense involving quantum computers, quantum networks using traditional telecom fibers, and the ability for a conventional computer to command and control a quantum device or network. Cyber security is intrinsically both national security policies, as well as a rapidly dynamic environment, which are all hard to qualify and quantify. The tipping point of debates on what constitutes an act of cyber war may occur in the form of a logical operation resulting in subsequent and devastating affects to human lives. Therefore the utmost care and precautions involving any research within the purview of quantum hacking cannot be stressed enough.

Lastly, the tactic of destroying physical systems with logical operations using procedures reliant on quantum technology with Gaussian techniques need be both defensible and undiscoverable. If under development or under consideration for development, the system need be conducted or deployed within an environment beyond the next-generation of cyber security systems, e.g. not only air-gapped. If unnecessary to use any computer system or electronic communication, the research must only be conducted using handwritten proofs, schematics, and communication until which a time may come to use such a device.

**Future Research Recommendations**

Remaining questions include to what degree a quantum computer can affect a conventional computer with next-generation defenses, as well as to what degree a quantum computer could affect physical systems. With the ability for logical operations, which may concretely influence the physical world, the development of both cyber and quantum weapons will only increase in focus for nation-states. The threat this poses to any opposing state by a malicious actor, be they another nation or a cyber vigilante, remains for mitigation. A recommended avenue for future research is not the development and deployment of such weapons, but rather the rapid development and deployment of defense systems against such attacks. Additionally, it remains how a conventional computer could exploit or attack a quantum computer.

If the current cryptographic system of RSA is vulnerable to fewer efforts less than acoustic hacking, the capabilities of a technique used by a nation-state operating a quantum computer pose valid levels of threat to security. In addition to this is analysis of trends and patterns within network communication of experimental protocols that reveal the full cipher of 4096-bit RSA encryption, the threat of a quantum computer under the command and control of malicious actors, regardless of affiliation, will prove devastating by virtue of principle.

Using discrete time as a Gaussian channel to create a vector of time, then using a complex algorithm for elliptic curves where the indices of imaginary components are a change in prime factorization without respect to entropy a new cryptographic system may be fully implementable rapidly. The use of a Gaussian channel curl by incorporating a Riemann surface as developed is a promising future direction of research as a reflective trapdoor within ECC.

## Conclusion

Radio waves may contain the ability to form a shield against attempts of intrusion, perhaps by a process of using the angular momentum to converge the propagation of waves as an eigenfunction upon ultraviolet light. The complex singularity of the radio wave has an amplitude equal to a resulting rate of coherence from Turchette and fellow researchers in their findings. By applying the rotation of $(\theta)$ as a value in a subsequent function of which the period is not $(\pi)$, a greater generation of a cyclic intersection between complex and real conjugates may aid in mitigation of quantum cracking. While there are artistic aspects of security that remain in the eye of the beholder, the concept of cyber security as a science is open for exploration and new frontiers.

While this work uses physics tied to number theory and components of artificial intelligence, perhaps the farthest reach conceivable result based on this work is to test the concepts of the orbital angular momentum as a trapdoor for recursion in the curl of a curve along an ellipse. Exploration of this requires the use of either an additive point without respect to the cofactor, or an eigenvector. The conceptualization of using logic bombs as mitigation is incomplete, though the implications and use of such systems appear to be a strong line of defense. Rejecting the idea that classical computers are incapable of executing quantum mechanics as an algorithmic implementation with the research conducted is feasible. Given classical computers operate using quantum mechanics the threat of a quantum computer for a malicious network attack or network exploit is unrealized, not impossible. Based on reflection, creating a hyperbolic surface tension curl to force decomposition of signals with radio waves, such that incoming malicious traffic undergoes reflective decomposition will aid in the defense against acoustic hacking.

.

# Appendices

## Appendix A – *QUINE*

Available for download at: https://github.com/FunctionAnalysis/qi-net/releases/tag/v1.0.1

```
:-op(1200,xf,~).
fact:device(input).
fact:device(udp).
fact:device(syn).
fact:device(ipa).
fact:device(port).
fact:(connected(input,port)):-
fact:(connected(port(2),computer2)).
fact:(connected(port(3),computer)):-
fact:(connected(port(4),computer)).
parse:connected(syn,udp,ipa):-parse:connected(syn,udp,syn),input(syn,udp,ipa).
parse:device(syn,udp,ipa).
parse:device(defines,classification,port).
parse:(output(classification(syn|X,udp|Y,ipa|Z))):-input(unknown(X,Y,Z)).
prolog:error_message(dde_error(Op,Msg)) -->
        [ 'DDE: ~w failed: ~w'-[Op,Msg] ].
unknown(output):-unknown(input).
classification(X):-(input(syn|[X])).
classification(unknown):-input(unknown).
classification(syn,udp,ipa):-unknown(input).
input(X,Y,Z):-port(input(X,Y,Z)).
input(X,Y,Z):-input(unknown(syn|X),(udp|Y),(ipa(Z))).
input(X,Y,Z):-parse:device(X,Y,Z).
input(X,Y,Z):-parse:connected(X,Y,Z).
input(Node,X,Y):-edge(X|Y,Node).
input(port):-fact:device(port).
input(unknown(classification(Y,Z,X))):-output(unknown(syn(X)),(udp(Y)),(ipa(Z))).
input(ipa):-unknown(input).
input(unknown(input)).
input(unknown):-unknown(input).
input(unknown(X,Y,Z)):-input(X,Y,Z).
output(X,Y,Z):-classification(X,Y,Z).
output(X,Y,Z):-(classification(X),(Y),(Z)).
matrix(node(A,B,C),edge([_]),bestf([],9999)):-matrix((node(A,B,C;d(_))),port(A),input(A)).
matrix(Line,Node,Distance):-edge(Line|Node+Distance).
matrix(A|Node_x;(B|Node1,(C|Node3)):-edge(A|Node1),edge(B|Node3), edge(C|Node_x)).
node(d([prime+1=prime])).
node(d([prime+2=prime])).
node(d([prime+1=prime])).
edge(X,Y):-(matrix(lattice,([])|X,Y)).
```

```
edge(X,Y):-fact:connected(X,Y).
edge([Node1,Node2];[(C;Node3)],[_]):-matrix(Node1|_,Node2|C,Node3).
edge([A,B];[B,C];[C,B]):-node(3),edge([A,B,C]),distance((node + edge = Distance)),matrix(edge,node,Distance).
edge((_;_;_)):-matrix((edge),node(2),node(3))).
edge([a]):-(number(prime),(edge([c]))).
edge([b]):-node(number(_)).
edge([c]):-node([prime1,prime2,prime3]|([a];[c];[b])).
distance(Prime):-
[(node(1),(Prime))]+[node(2),(Prime)]+[node(3),(Prime)]=(node(1+2=2),node(2+3=2),node(1+3=4),edge(3)).
'$dde_disconnect'(ipa(Service, Topic, _Self)) :-
        dde_service(Service, Topic, _, _, _, _).
        '$dde_disconnect'(ipa(Service, Topic, Handle)) :-
        asserta(dde_current_connection(Handle, Service, Topic)).
        '$dde_disconnect'(ipa).
'$dde_disconnect'(Handle) :-
        retractall(dde_current_connection(Handle, _, _)).
port(_) :-
        strip_module(port((Module)--> Plain),Module,Plain),
        Plain =.. [Vuln|Args],
        gather_args(Args, Values),
        Goal =.. [Vuln|Values],
        Module:Goal,
        port(port->close).
port(close):-(rl_write_history(port)).
port(classification(on_signal(Vuln|Scan,Vuln|Open,Open))):-(parse:output(Scan)).
port(retractall(Vuln)):-port(Vuln).
port(retractall(parse:parse(Vuln))):-port(Vuln).
port(Open|Scan):-('$dde_execute'((port(_)),Scan,Open)).
((port(Access;Open)):-('$dde_request'(((Access)),write([vulnerabilities]),(Open),(port(_))))).
(((port(IP)) :-
        dde_current_connection((Scan|Vuln),Scan, Vuln),IP)).
port((_,_)):-'$dde_disconnect'((_,_,_,_)).
gather_args([], []).
gather_args([+H0|T0], [H|T]) :- !,
        unknown(port(H0, H)),
        gather_args(T0, T).
gather_args([H|T0], [H|T]) :-
        gather_args(T0, T).
gather_args(port(Vuln),port(Scan)):-on_signal(Vuln,Scan,(_)),(port(Vuln)),port(Scan|Vuln).
gather_args(file(Mode, Title), File) :-
        '$append'(Filter, [tuple('All files', '*.*')], AllTuples),
        Filter =.. [chain|AllTuples],
        current_prolog_flag(hwnd, HWND),
        working_directory(CWD, CWD),
```

```prolog
        call(get(@display, win_file_name,
                Mode, Filter, Title,
                directory := CWD,
                owner := HWND,
                File)).
rl_write_history(port):-rl_read_history(port).
'$dde_request'(syn, port(Vuln), ipa(Vuln), udp).
'$dde_request'(Handle, Topic, Item, Answer) :-
        dde_current_connection(Handle, Service, Topic),
        dde_service(Service, Topic, Item, Value, Module, Goal), !,
        Module:Goal,
        Answer = Value.
'$dde_request'(_Handle, Topic, _Item, _Answer) :-
        throw(error(existence_error(dde_topic, Topic), _)).
'$dde_request'(Service, Topic, _Self,Vuln) :-
        dde_service(Service, Topic, _, _,Vuln, _).
'$dde_request'((Vuln|Scan),Vuln,Open, (_)):-(dde_current_connection(Scan,Vuln,Open)).
'$dde_request'(Handle, Topic, Item, Answer) :-
        dde_current_connection(Handle, Service, Topic),
        dde_service(Service, Topic, Item, Vuln, port, close(Vuln)), !,Answer = close.
'$dde_request'(_Handle, Topic, _Item, _Answer) :-
        throw(error(existence_error(dde_topic, Topic), _)).
'$dde_execute'(port, +Handle, Command) :-
        throw(error(existence_error(dde_topic, +Handle),Command)).
'$dde_execute'(port(Vuln),write([vulnerabilities]),(command|(port(Vuln)))).
'$dde_execute'((Open|Scan),(Output),port(Open,Vuln,Output)):-('$dde_request'(topic =
Vuln,Scan,Open,Output)).
'$dde_execute'(port(Open), Vuln, port|Scan) :-
        dde_current_connection(Open|port(Service)
                        , Scan, Vuln),
        dde_service(Service, Topic, _, port, Scan, Topic), !, port(Topic|Vuln).
'$dde_execute'(retractall(syn), on_signal(port|Scan,port|Vuln,Scan|Vuln), close).
'$dde_execute'(Handle, Topic, Command) :-
        dde_current_connection(Handle, Service, Topic),
        dde_service(Service, Topic, _, Command, Module, Goal), !,
        Module:Goal.
'$dde_execute'(_Handle, Topic, _Command) :-
        throw(error(existence_error(dde_topic, Topic), _)).
(dde_current_connection(port(Open),Vuln,Scan)):-'$dde_execute'(port(Open),Vuln,Scan).
((dde_service(Scan, _, _, _, ([_]),(_))):-(port(Scan))).
prolog:error_message(dde_error(Op,Msg)) -->
[ 'DDE: ~w failed: ~w'-[Op,Msg] ].
~(_):-not(_).
~(P):-!,(fail),not(P);true.
f( l(_,F/_),F).
f( t(_,F/_,_),F).
```

```
h(ipa,syn).
s(ipa,syn,udp).
t(N,F/G,Sub):-l(N,F/G,Sub).
l(N,F/G,Sub):-(t(N,F/G,Sub)).
bagof(syn/ipa).
goal(_):-goal(n).
bestf(Vuln,Solution):-
        expand(Vuln,l(Vuln,0/0),9999,_,yes,Solution).
bestf([T|_],F):-
        f(T,F).
bestf([],9999).
expand(P,l(N,_),_,_,yes,[N|P]):-goal(N).
expand(P,Tree,Bound,Tree1,Solved,Solution):-port(P),port(Tree|Bound|Tree1;Solved|Solution).
expand(P,l(N,_),_,_,yes,[N|P]):-goal(N).
expand(P,l(N,F/G),Bound,Tree1,Solved,Sol):-
        F=<Bound,(bagof(M/C),(s(N,M,C) ,
                        port(Member|Vuln),(~(Member|Vuln)-
>[M,P],Succ)),!,succlist(G,Succ,Ts),bestf(Ts,Fl),
                expand(P,t(N,Fl/G,Ts),Bound,Tree1,Solved,Sol);Solved=0).
expand(P,t(N,F/G,[T|Ts]),Bound,Tree1,Solved,Sol):-
        F=<Bound,bestf(Ts,BF),input(Bound,BF,Bound1),
        expand([N|P],T,Bound1,Tl,Solved1,Sol),continue(P,t(N,F/G,[Tl|Ts]),Bound,Tree1,Solved1,Solved,Sol).
expand(_,t(_,_,[]),_,_,never,_):-!.
expand(_,Tree,Bound,Tree,no,_):-f(Tree,F),F>Bound.
continue(_, _, _, yes, yes, open,_).
continue( P, t(N, Fl/G, [Tl|Ts]), Bound, Tree1, Solved, Sol,_):-
        insert(Tl, Ts, NTs),
        bestf(NTs,Fl),
        expand(P, t(N, Fl/G, NTs), Bound, Tree1, Solved,Sol).
succlist(_, [], []).
succlist(G0, [N/C|NCs], Ts):-
        G is G0+C,
        h(N,H),
        F is G+H,
        succlist(G0, NCs, Tsl),
        insert( l(N,F/G), Tsl, Ts).
insert(T,Ts,[T|Ts]):-
        f(T,F),bestf(Ts,Fl),
        F=<Fl,!.
insert(T,[Tl|Ts],[Tl|Tsl]):-
        insert(T,Ts,Tsl).
```

## Appendix B – Supplemental Proofs

### Unique Existence of Complex Analytic Function($\varphi$):

**Theorem**: There exists some function $(f(x) = \varphi)$, and $\{(f(z) = \varphi) \to [(\lfloor \varphi \rfloor \in z)]\}$

**Lemma**: There exists some constant K where $(K \neq \infty)$, and $((K = 0) \in (\theta|_K))$

$$\left\{\left((2K\pi + z) \uparrow \left(\frac{z}{\pi}\right)\right) \uparrow \overrightarrow{(x, y, z)}\right\}, \text{ where } \left[4\theta \sum_{K=0}^{\infty} \frac{(-1)^K \left((2K\pi + x)\left(\frac{x}{\pi}\right)_K^3\right)}{(K!)^3}\right]$$

**Proof**

$$\left[\lfloor \varphi \rfloor \to (ie^{-i\theta} - ie^{i\theta}), (f(z) = 2\theta^2)\right] \to (K = 0)$$

Therefore $\{(f(z) = \varphi) \to [(\lfloor \varphi \rfloor \in z)]\}$ where $(K = 0)$. If $\left\{\lceil \varphi \rceil \uparrow \left(\lim_{K \to \infty} \varphi \cong 4\theta\right)\right\}$ where:

$[((4\theta) \cong 179.21°) \uparrow (\theta|_K)]$. Therefore, $(\varphi) = \left[4\theta \sum_{K=0}^{\infty} \frac{(-1)^K \left((2K\pi + z)\left(\frac{z}{\pi}\right)_K^3\right)}{(K!)^3}\right]$ where

$$(\varphi) = \left[4 * (179.21°|_K) \sum_{K=0}^{\infty} \frac{-1\left(\left(\frac{z}{1}\right)\left(\frac{z}{\pi}\right)_0^3\right)}{1}\right]$$

Then

$$4 * (179.21°|_K) \sum_{K=0}^{\infty} -\left(\left(\frac{z}{1}\right)\left(\frac{z}{\pi}\right)_0^3\right) = \left[4 * (179.21°|_K \sum_{K=0}^{\infty} -\left(\frac{z^4}{\pi^3}\right)\right]$$

Where $\left[\left(\frac{z^4}{\pi^3}\right) = w\right]$, and $\left\{\lceil \varphi \rceil = \left(12.511 - \left(\frac{z^4}{\pi^3}\right)\right)\right\}$

## ℌ Condition Satisfiability

**Theorem**: x greater than zero is a time like vector and x less than zero is a space like vector. If x equals zero it is a null vector or light like. With this system, x represents the space-time vectors.
$(x \neq 0, x^2 \cong 1)$, and $\{0^{-i\pi n} - 0^{i\pi n}\} = x$.

**Lemma**:

$$\forall (x,y) \in \mathfrak{H}, (x,y) > 0, x \neq 0$$

$$\forall (x,y) \in \mathfrak{H}, ((x,y) = (x,y))$$

$$\forall (x,y) \in \mathfrak{H}, (\alpha x, y) = (x,y)$$

$$\forall (x,y) \in \mathfrak{H}, [(x+y, z) = (x,z) + (y,z)]$$

**Proof**: When $\{x = \{0^{-i\pi n} - 0^{i\pi n}\}\}$ $(n)$ must be shown not to reduce the value of $(x)$ to zero. Let $(n = 0)$. If $(n = 0 \to x = 0)$.

$\forall x \{(x \neq 0) \to (x = 0^{-i\pi n} - 0^{i\pi n})\} \, iff \, (x^2 \cong 1)$.

If $(0^{-i\pi n} - 0^{i\pi n})^2 \neq 0$, then $(x^2 \cong 1)$.

$(0^{-i\pi n} - 0^{i\pi n})^2, n = 0, \{0^{-2i\pi n} + 0^{i\pi n} - 2\}$ remains indeterminate and has odd parity. No solutions where $(0^{-i\pi n} - 0^{i\pi n}) = 0$ exists.

$$\therefore x \neq 0$$

**Cyclic Collinear Group:**

**Theorem:** $(G)$ is a homomorphic cyclic group of which $(H)$ and $(G)$ such that:

$$\left\{\left(\frac{G}{H}\right), (y_i mod(H)), (y_i \neq G)\right\}.$$

**Lemma:** Let Z be an element of H and K, $[Z \in (H + K)]$ and G as a cyclic function of an element of Z, $[G = \{\hat{a}: (n \in Z)\}]$

$$(G = \langle a \rangle)$$

**Proof:** Let $[\langle a \rangle \in \{G\}]$ where $(a)$ is $(\hat{a}: (n \in Z))$. Given $[Z \in (H + K)]$. Such that when $(a \uparrow p)$ are in series, $(a_0 + a_1 p \dots a_n p_n^n)$ and $(x_i = p_n^n)$, Then:

$$\left\{(x_i \in K) \to \left((H \oplus x_i) \uparrow (Z \in (K + H))\right)\right\} \text{ And } \{(\langle a \rangle \leq G) = (\hat{a}: (n \in Z))\} \text{ where}$$

$(G = \sum a_i y_i)$ Moreover $[p_n^n = Z - \sum a_i x_i]$, then when $\left\{t^* = \left[G \uparrow \left[p_n^n \oplus \left(\aleph_0 \in \frac{G}{H}\right)\right]\right]\right\}$ resulting in $(y_i = G)$.

If $(y_i = G)(x_i \in Z), (f(x_i) = \hat{a}) \left\{\left(\frac{y_i}{H}\right) = (y_i mod(H))\right\}$ and if true:

$\left\{\left(\frac{G}{H}\right) = (y_i mod(H))\right\}$, But then $(G = (H \oplus K)), iff (G = K)$,

$$\therefore \left\{\left(\frac{G}{H}\right), (y_i mod(H)), (y_i \neq G)\right\}$$

∎